\documentclass{article}

\PassOptionsToPackage{authoryear, sort&compress}{natbib}

\usepackage[final]{neurips_2024}
\usepackage[many]{tcolorbox}
\usepackage[edges]{forest}
\usepackage{array}

\newcolumntype{C}{>{\centering\arraybackslash}X}
\usepackage[colorinlistoftodos,prependcaption,textsize=tiny]{todonotes}
\usepackage{xargs} 
\usepackage{wrapfig}
\usepackage{threeparttable}
\usepackage{enumitem}
\usepackage{makecell}
\usepackage[utf8]{inputenc} %
\usepackage[T1]{fontenc}    %
\usepackage{url}            %
\usepackage{booktabs}       %
\usepackage{amsfonts}       %
\usepackage{nicefrac}       %
\usepackage{xcolor}         %
\usepackage{microtype}
\usepackage{graphicx}
\usepackage{floatrow}
\usepackage{subcaption}
\usepackage{listings}
\usepackage{longtable}
\usepackage{array}
\usepackage{tcolorbox} %
\usepackage{multirow}
\usepackage{marvosym} 
\usepackage{forest}
\usepackage[bookmarks=true,colorlinks=true,linkcolor=black,urlcolor=magenta,citecolor=green!50!black]{hyperref}

\usepackage{tabularx}
\usepackage{pifont}

\usepackage{amsmath}
\usepackage{amssymb}
\usepackage{mathtools}
\usepackage{amsthm}
\theoremstyle{plain}

\theoremstyle{definition}

\theoremstyle{remark}

\usepackage{bm} 
\usepackage{algorithm, algorithmic}

\newcommandx{\yhr}[2][1=]{\todo[linecolor=blue,backgroundcolor=blue!25,bordercolor=blue,#1]{hr: #2}}

\usepackage{etoolbox}
\usepackage{natbib}
\newcounter{bibcount}
\makeatletter
\patchcmd{\@lbibitem}{\item[}{\item[\hfil\hspace{2.5em}\stepcounter{bibcount}{[\thebibcount]}\hspace{0.em}}{}{}
\makeatother
\usepackage[nameinlink,capitalise]{cleveref}
\crefname{section}{§}{§§}
\Crefname{section}{§}{§§}
\crefname{lemma}{lemma}{lemmas}
\Crefname{lemma}{Lemma}{Lemmas}
\crefname{thm}{theorem}{theorems}
\Crefname{thm}{Theorem}{Theorems}

\definecolor{TrendBlue}{RGB}{24, 88, 140}    
\definecolor{TrendLight}{RGB}{248, 250, 252} 

\newtcolorbox{trendminimal}[1][New Trend]{
    enhanced,
    arc=2pt,                 
    boxrule=0.5pt,           
    colframe=gray!25!white,  
    colback=TrendLight,      
    borderline west={3.5pt}{0pt}{TrendBlue}, 
    coltitle=TrendBlue,      
    fonttitle=\bfseries\sffamily, 
    title={#1},
    left=8pt, right=8pt, top=4pt, bottom=4pt,
    boxsep=2pt,
}

\newtcolorbox{takeaway}[1][New Trend]{
    enhanced,
    colback=gray!3!white,
    colframe=gray!30!white,
    boxrule=0.5pt,
    arc=3pt,
    drop fuzzy shadow=black!8!white, 
    title={#1},
    fonttitle=\bfseries\sffamily,
    coltitle=TrendBlue,
    attach boxed title to top left={xshift=12pt, yshift=-\tcboxedtitleheight/2},
    boxed title style={
        colback=white,       
        colframe=gray!30!white,
        boxrule=0.5pt,
        arc=2pt,
        left=5pt, right=5pt, top=1pt, bottom=1pt
    },
    left=8pt, right=8pt, top=12pt, bottom=6pt,
}

\title{Retrieval-Augmented Code Generation: A Survey with Focus on Repository-Level Approaches}

\author{%
  Yicheng Tao\textsuperscript{1, *}, Yuante Li\textsuperscript{1, *}, Yao Qin\textsuperscript{2}, Yepang Liu\textsuperscript{3}
  \\[0.1cm]
\textsuperscript{1}Carnegie Mellon University\\
\textsuperscript{2}Chinese University of Hong Kong\\
\textsuperscript{3}Southern University of Science and Technology\\
[0.1cm]
\small \texttt{yichengtao@cmu.edu} \quad \texttt{yuantel@cs.cmu.edu} \\ \small  \texttt{1155240806@link.cuhk.edu.hk}
\quad \texttt{liuyp1@sustech.edu.cn} \\
[0.1cm]
\small \textsuperscript{*}Equal Contribution.
}

\begin{document}

\newtcolorbox{definitionbox}{
  colback=blue!5!white,    %
  colframe=blue!75!black,  %
  boxrule=0.8pt,           %
  arc=4pt,                 %
  left=6pt, right=6pt, top=6pt, bottom=6pt, %
  fonttitle=\bfseries,
  before skip=10pt, after skip=10pt
}

\maketitle

\begin{abstract}

Recent advances in large language models (LLMs) have significantly improved automated code generation. While existing approaches have achieved strong performance at the function and file levels, real-world software engineering requires reasoning over entire repositories, including cross-file dependencies, evolving execution environments, and global semantic consistency. This challenge has led to the emergence of Repository-Level Code Generation (RLCG), where models must retrieve, organize, and utilize repository-scale context to generate coherent and executable code changes.
To address these challenges, Retrieval-Augmented Generation (RAG) has become an increasingly important paradigm for repository-level code intelligence. In this survey, we present a comprehensive review of Retrieval-Augmented Code Generation (RACG), with a particular focus on repository-level approaches. Rather than viewing RACG as a static ``retrieve-then-generate'' pipeline, we characterize it as a coupled and evolving process involving context construction, retrieval optimization, generation, and environment interaction.
We organize existing methods through a unified analytical framework spanning retrieval substrate, control regime, and evaluation setting. Based on this framework, we systematically examine retrieval strategies, graph-based and non-graph-based retrieval paradigms, training-driven optimizations, and autonomous agent architectures. We further summarize widely used datasets, benchmarks, and system configurations, and discuss key challenges including scalability, reliability, efficiency, and the necessity boundary between RACG and long-context LLMs. Through this survey, we aim to provide a structured understanding of the rapidly evolving RACG landscape and highlight promising directions for future AI-powered software engineering research.

\end{abstract}

\begin{figure}[h]
    \centering
    \includegraphics[width=0.3\textwidth]{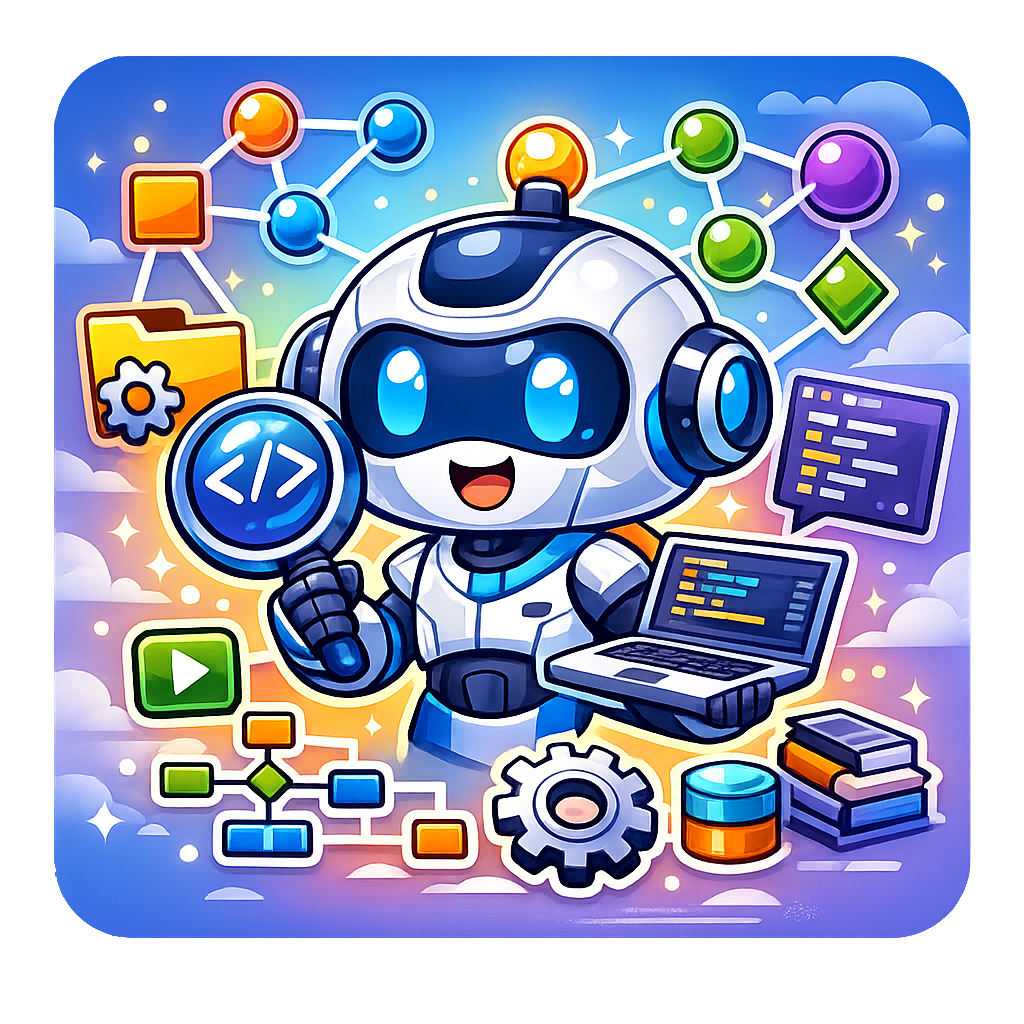}
\end{figure}
\vspace{0.3cm}

\clearpage

{\color{black}
\tableofcontents
}

\clearpage

\hypersetup{linkcolor=red}

\section{Introduction}\label{sec: introduction}

\begin{quote}
    "If, as we confront some task, a part of the world functions as a process which, were it done in the head, we would have no hesitation in recognizing as part of the cognitive process, then that part of the world is part of the cognitive process." \citep{97be971b-f829-3ce8-80b8-a2e1dc1e11df}
\end{quote}

\begin{figure}
    \centering
    \includegraphics[width=0.75\linewidth]{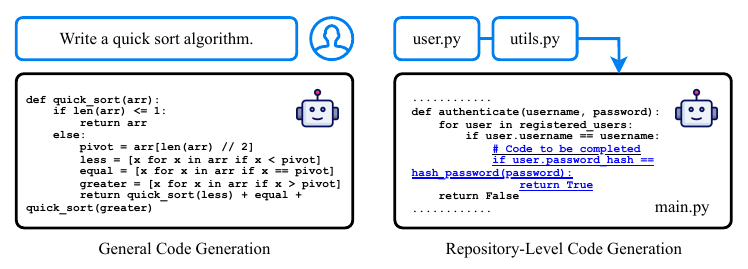}
    \caption{Comparison between General Code Generation and Repository-Level Code Generation}
    \label{fig:compare}
    \vspace{-10pt}
\end{figure}

Recent advances in large language models (LLMs) are transforming software development from code suggestion to broader software engineering capabilities~\citep{jimenez2024swebench}. Despite the inherent complexity of real-world development, characterized by large codebases, cross-file dependencies, and diverse tooling, agentic systems such as OpenClaw\footnote{\url{https://github.com/OpenClaw/OpenClaw}}, Claude Code\footnote{\url{https://www.anthropic.com/claude-code}} and Codex\footnote{\url{https://openai.com/index/openai-codex/}} demonstrate that LLMs can operate effectively in these environments and make progress. These advances mark a shift from isolated code generation to more integrated and practical software engineering workflows, as shown in Figure~\ref{fig:compare}.
In practice, development is repository-centric, where functionality arises from interactions across files. Consequently, models must not only produce locally correct code, but also maintain global consistency and adapt to an evolving execution environment. These requirements define the core challenges of modern code generation.



\paragraph{Challenges.}
The first challenge is \textit{long-range dependency modeling}: critical information is distributed across files, documents, and dependencies, while changes in one file may affect distant modules, making it difficult to identify minimal but sufficient context and maintain cross-file consistency~\citep{assogba2025evaluating}. 
The second is the \textit{efficiency--correctness trade-off}: issue resolution prioritizes accuracy, whereas real-time completion demands low latency. This trade-off is further complicated by model limitations—strong models may memorize common libraries, while weaker ones suffer from context overload~\citep{wang-etal-2025-coderag}. Scaling model size or context length only partially alleviates these issues and is often impractical for local deployment. 
Finally, \textit{practical constraints} further limit applicability: privacy and compliance restrict cloud-based use of proprietary code~\citep{llm_privacy_surveys, local_llm_privacy_benefits}, and most LLMs are trained in outdated public repositories~\citep{otterly2024knowledge}, reducing adaptability in real-world and enterprise settings.

The above challenges indicate that, in real-world settings, code generation is no longer a simple problem, but a complex process involving context selection and external information acquisition. In this context, Retrieval-Augmented Generation (RAG) offers an effective way to address context limitations by dynamically incorporating external knowledge during reasoning. In the code domain, this paradigm has evolved into \textbf{\underline{R}etrieval-\underline{A}ugmented \underline{C}ode \underline{G}eneration (RACG)}, which enhances generation quality by retrieving relevant code and structural information.

However, in repository-level scenarios, RACG is no longer a simple two-stage ``retrieve-then-generate'' pipeline. Instead, it becomes a dynamic process that can be coupled with customized indexing and structure-aware retrieval. We characterize RACG as a unified process in which the model (i) constructs and retrieves context from large-scale code repositories, and (ii) generates or modifies code based on the retrieved information and current system state. 
In addition to this core process, various enhancements—such as iterative refinement, multi-source context acquisition, and tool-assisted reasoning—can be incorporated to improve performance~\citep{repocoder,evor,ma2025toolintegratedreinforcementlearningrepo}. This formulation naturally aligns with \textbf{\underline{R}epository-\underline{L}evel \underline{C}ode \underline{G}eneration (RLCG)}, which represents a new and challenging problem.

Existing surveys have reviewed general RAG systems~\citep{gao2024retrievalaugmentedgenerationlargelanguage,surveyrag}, LLM-based code generation~\citep{surveycodegeneration,dong2025surveycodegenerationllmbased}, and code generation agents for software engineering~\citep{surveycodeagent}. However, these works do not systematically examine how retrieval is used to construct, select, and integrate the repository-level context. This survey fills this gap by providing a systematic overview of Retrieval-Augmented Code Generation in repository-level scenarios, organized around three questions: (i) how RACG methods are designed; (ii) how they are formulated and evaluated; and (iii) concerns and necessity boundary. Table~\ref{tab:survey_comparison} summarizes the positioning of this survey relative to previous work. 

\begin{table}[h]
\centering
\caption{Comparison with related surveys. \checkmark~= fully covered, $\sim$~= partially covered, $\times$~= not covered.}
\label{tab:survey_comparison}
\small
\begin{tabular}{lcccc}
\toprule
\textbf{Survey} & \textbf{RAG Focus} & \textbf{Repo-Level} & \textbf{Code Gen} & \textbf{Agent Arch.} \\
\midrule
Gao et al.~\citep{gao2024retrievalaugmentedgenerationlargelanguage} & \checkmark & $\times$ & $\times$ & $\times$ \\
Fan et al.~\citep{surveyrag} & \checkmark & $\times$ & $\times$ & $\times$ \\
Jiang et al.~\citep{surveycodegeneration} & $\times$ & $\times$ & \checkmark & $\times$ \\
Dong et al.~\citep{dong2025surveycodegenerationllmbased} & $\sim$ & $\times$ & \checkmark & \checkmark \\
He et al.~\citep{surveycodeagent} & $\times$ & $\times$ & $\sim$ & \checkmark \\
\textbf{This survey} & \checkmark & \checkmark & \checkmark & \checkmark \\
\bottomrule
\end{tabular}
\end{table}

\paragraph{Key Contributions.}
\begin{itemize}[leftmargin=2em]
    \item We establish Repository-Level Code Generation as a unified problem formulation, explicitly capturing cross-file dependencies, dynamic context, and system-level constraints.
    \item We provide the first survey of Retrieval-Augmented Code Generation, reframing RAG as a coupled, evolving process rather than a static pipeline.
    \item We introduce a unified analytical framework that organizes RACG along retrieval, control, and integration dimensions, covering both model design and agentic workflows.
    \item We systematically evaluate existing methods, benchmarks, and system configurations, revealing key trade-offs between efficiency, scalability, and correctness.
    \item We identify critical gaps in current research and outline future directions toward better RACG systems.
\end{itemize}
\section{Preliminary and Methodological Foundations}\label{sec: preliminaries}

This section formalizes the key concepts underlying Retrieval-Augmented Code Generation in repository-level settings. We focus on minimal yet sufficient abstractions that align with the paradigm introduced in Section~\ref{sec: introduction}. RLCG defines the task setting, while RACG describes a family of retrieval-augmented methodologies to solve repository-level code generation problems.

\subsection{Repository-Level Code Generation (RLCG)}
\label{rlcg-def}
Repository-Level Code Generation (RLCG) refers to the generation or modification of code in the context of an entire software repository, rather than isolated functions or files. 

Formally, let a repository be denoted as $\mathcal{R} = \{f_1, f_2, \dots, f_n\}$, where each $f_i$ represents a file containing code, documentation, or configuration. Given a task specification $q$ (e.g., a bug report, feature request, unfinished code snippet), the objective of RLCG is to produce a modification $\Delta \mathcal{R}$ such that:
\begin{equation}
\mathcal{R}' = \mathcal{R} \oplus \Delta \mathcal{R}
\end{equation}
satisfies both local correctness (syntactic and semantic validity) and global consistency across the repository.

Typical applications include
\begin{itemize}[leftmargin=2em]
    \item \textit{Cross-file Code Completion:} Predicting or synthesizing missing code segments by leveraging repository-wide context, from single lines to full functions.
    \item \textit{Issue or PR Resolution:} Automatically fixing reported problems by locating relevant files and generating patches consistent with project conventions.
\end{itemize}
We discuss a broader range of applications in Section~\ref{sec: tasks}.

Unlike function-level generation, RLCG requires reasoning over long-range dependencies, cross-file interactions, and system-level constraints. The generation process is inherently state-dependent, as intermediate changes may alter the repository structure and influence subsequent decisions.

\subsection{Retrieval-Augmented Code Generation (RACG)}

\begin{figure}
    \centering
    \includegraphics[width=0.9\linewidth]{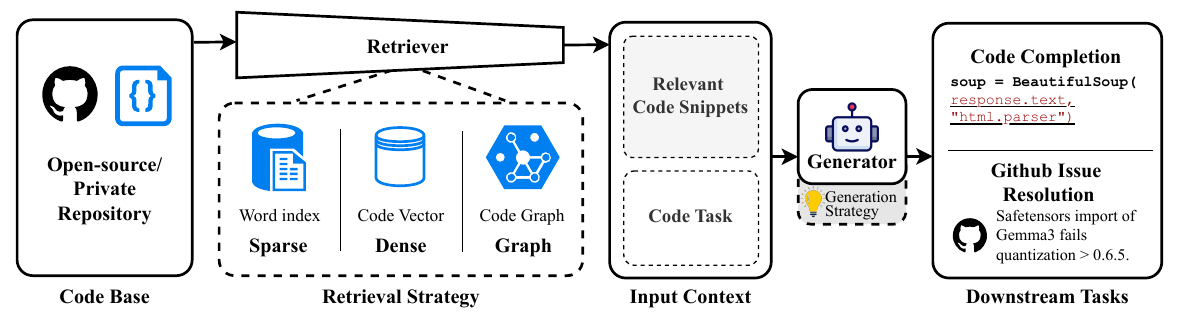}
    \caption{Retrieval-Augmented Code Generation, focusing on Repository-Level approaches.}
    \label{fig:rlcg}
    \vspace{-10pt}
\end{figure}

Retrieval-Augmented Code Generation enhances LLMs with external software knowledge via retrieval. Unlike pure LLM generation, RACG dynamically retrieves relevant information such as code, documentation, or graphs to construct context-aware prompts.

In its basic form, RACG (Figure \ref{fig:rlcg}) consists of two components: (i) \textbf{Retriever} $\mathcal{M}_r$, which selects the relevant context $C \subseteq \mathcal{R}$ given a query or intermediate state. It typically operates through several structured strategies, including \textit{identifier matching} based on exact names or signatures, \textit{sparse retrieval} using lexical or sparse-vector methods such as TF-IDF~\citep{manning2008introduction} or BM25~\citep{10.1007/978-1-4471-2099-5_24}, \textit{dense retrieval} using neural embeddings such as CodeBERT~\citep{codebert} or UniXcoder~\citep{guo-etal-2022-unixcoder} with nearest-neighbor search, \textit{graph-based retrieval} leveraging structures such as abstract syntax trees and call graphs via traversal or subgraph matching, and \textit{hybrid retrieval} that integrates lexical, embedding, and structural signals to balance precision and recall. (ii) \textbf{Generator} $\mathcal{M}_g$ is a language model such as GPT-5.5~\citep{openai2025gpt5} or CodeLlama~\citep{codellama} that produces code conditioned on the context retrieved.

Traditional RAG follows a static pipeline:
\begin{equation}
C = \mathcal{M}_r(q), \quad y = \mathcal{M}_g(q, C)
\end{equation}

However, in repository-level settings, RACG departs from this simple paradigm. Instead of treating retrieval as a static preprocessing step, RACG operates as a coupled process where retrieval and generation are interdependent and evolve over time. Recent work also explores iterative or agent-style RACG frameworks~\citep{repocoder,orcaloca}. 

\section{Methodology}\label{sec: method}

This section outlines the methodology used to conduct a systematic literature review.
We adopt a structured review approach inspired by established guidelines in software engineering literature~\citep{10.5555/2227115.2227123,surveycodeagent,surveycodegeneration}.
The methodology encompasses research question formulation, data collection, inclusion and exclusion filtering, quality assessment, snowballing, and topic categorization.

\begin{itemize}[leftmargin=2em]
    \item[\ding{224}] \textbf{RQ1: How does RACG evolve from simple retrieval paradigms to complex autonomous architectures?} \\
    This question investigates the methodological shift from semantic retrieval to structure-aware graph retrieval. It further explores how training-based optimizations and agentic architectures address complex task resolution.
    
    \item[\ding{224}] \textbf{RQ2: What are the core configurations, task environments, and evaluation practices for RACG systems?} \\
    This question examines the ecosystem and implementation of RACG, covering the impact of diverse data sources and programming language paradigms. It also evaluates how downstream tasks and current benchmark metrics reflect the demands of practical software engineering.
    
    \item[\ding{224}] \textbf{RQ3: What are the concerns of RACG systems and where is its necessity boundary?} \\
This question analyzes key system-level bottlenecks, including reliability, scalability, and structural complexity. 
It further examines when retrieval remains necessary versus when long-context modeling is sufficient, and outlines directions for bridging this gap toward real-world deployment.
\end{itemize}

\subsection{Data Collection Process}

To address our research questions and construct a comprehensive corpus on RACG, we adopted a systematic two-stage data collection pipeline, with the primary focus placed on \textbf{methodological advances }in retrieval-augmented and repository-level code generation.

In the first stage, we determined our core search keywords by analyzing accepted papers related to this topic. This initial step allowed us to derive and refine a precise set of search terms strongly associated with RACG. 

In the second stage, we collected all publications from leading artificial intelligence and software engineering venues and journals (\texttt{ICLR}, \texttt{NeurIPS}, \texttt{ICML}, \texttt{IJCAI}, \texttt{AAAI}, \texttt{JMLR}, \texttt{ACL}, \texttt{EMNLP}, \texttt{ICSE}, \texttt{FSE}, \texttt{ASE}, \texttt{TOSEM}, \texttt{TSE}, and \texttt{ISSTA}). We then conducted keyword similarity searches across multiple bibliographic platforms, including \texttt{ACM Digital Library}, \texttt{IEEE Xplore}, \texttt{arXiv}, \texttt{OpenReview}, and the \texttt{ACL Anthology}. For journals and platforms with anti-crawling mechanisms, we supplemented automated retrieval with manual scanning. For preprints on \texttt{arXiv}, we implemented a hybrid approach: we first queried the arXiv API using the keywords, and subsequently applied similarity filtering to isolate the most relevant papers.

The search period spanned from January~1,~2023 to March~31,~2026. This process resulted in 243 candidate papers from the targeted conferences and journals, along with 610 papers from \texttt{arXiv}. After consolidating these sources and removing duplicates, the resulting candidate pool was screened according to the criteria detailed in Section~\ref{inclusionandexclusioncriteria}.
\subsection{Inclusion and Exclusion Criteria}
\label{inclusionandexclusioncriteria}

A paper was included if it met all the following:
\begin{itemize}[leftmargin=2em]
    \item The task involves retrieval-augmented, cross-file, or repository-level settings that enhance code generation.
    \item The study proposes a novel RAG-based approach or analyzes the impact of retrieval on generation quality.
    \item The full text is accessible and written in English.
\end{itemize}

A paper was excluded if it:
\begin{itemize}[leftmargin=2em]
    \item Focuses solely on standalone code generation or static analysis tools without pretrained Language Models.
    \item Uses RAG only as a minor component without methodological contribution.
  \item Target tasks such as vulnerability or clone detection, or purely retrieval model training without generation (except for issue localization tasks, which are retained despite being purely retrieval-based due to their high relevance to repository-level settings).
    \item Presents only conceptual discussions or ethical commentary without technical contributions.
\end{itemize}


\subsection{Quality Assessment and Final Corpus}

To ensure the rigor and completeness of our review, each candidate article was evaluated for: (1) relevance to RACG, (2) methodological clarity, and (3) reproducibility. Works that did not meet these criteria were excluded. For preprints sourced from \texttt{arXiv}, we additionally required a detailed method section and solid experimental results to compensate for the lack of formal peer review. 

Furthermore, backward snowballing was applied to the reference lists of included papers to identify foundational and related works potentially missed by the initial keyword searches. 

After completing these rigorous screening and quality checks, a final corpus of 152 papers was retained, comprehensively covering the intersections of software engineering, natural language processing, and machine learning.

\subsection{Topic Categorization and Analysis}

We manually annotated each paper along multiple dimensions, including task type, system architecture, retrieval method, evaluation setting, and target domain. These annotations support both the quantitative distribution analysis and the taxonomy developed in the following sections.

\begin{figure}[t]

    \centering

    \includegraphics[width=0.72\linewidth]{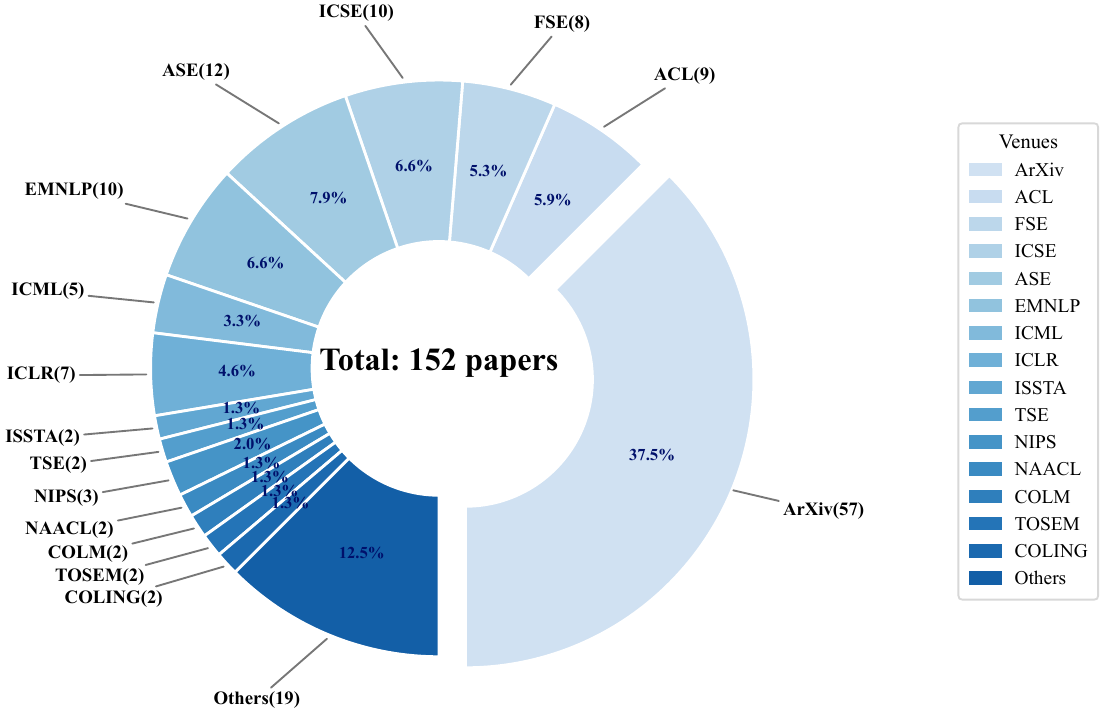}

    \caption{Distribution of selected RACG papers across publication venues.}

    \label{fig:admission-distribution}

\end{figure}

Figure~\ref{fig:admission-distribution} shows the publication venue distribution of the 152 selected papers. \texttt{arXiv} constitutes the largest category, with 57 papers, followed by major software engineering and NLP venues such as \texttt{ASE}, \texttt{ICSE}, \texttt{EMNLP}, \texttt{ACL}, and \texttt{FSE}. This distribution reflects the rapid iteration cycle of RACG research, where many works first appear as preprints while increasingly entering established venues in software engineering, natural language processing, and machine learning.

\begin{figure}[t]

    \centering

    \includegraphics[width=0.8\linewidth]{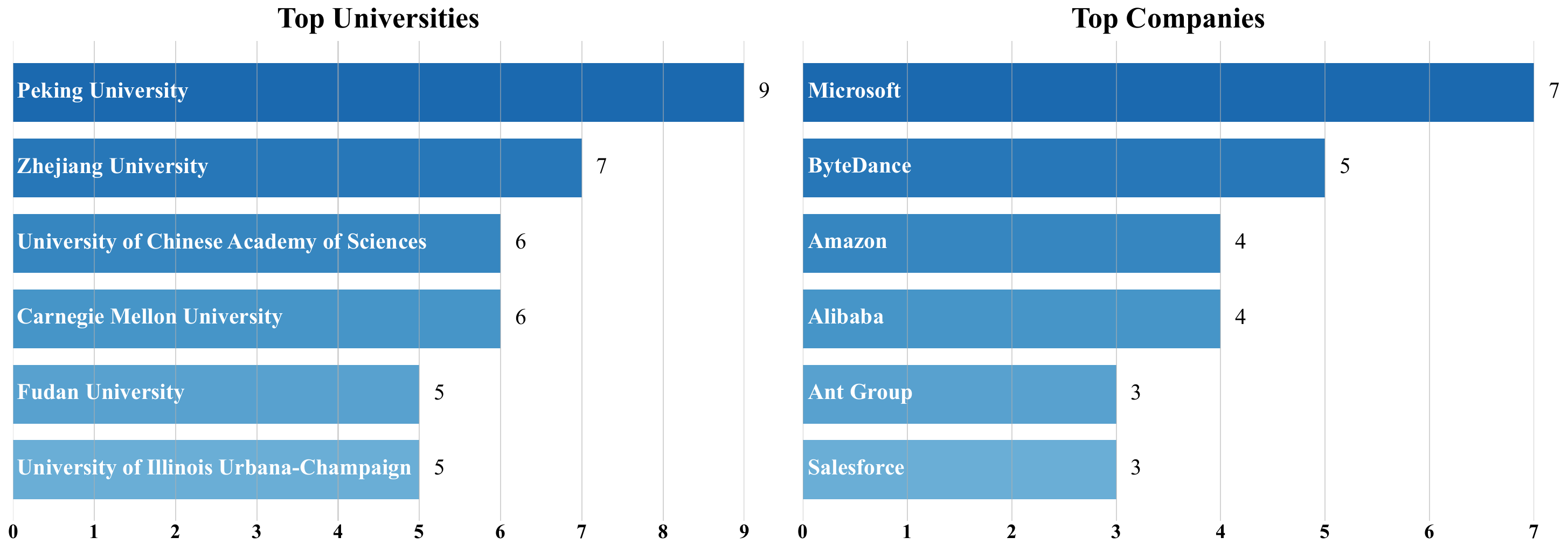}

    \caption{Top contributing universities and companies in RACG research.}

    \label{fig:institutions-dual}

\end{figure}

Figure~\ref{fig:institutions-dual} further summarizes the leading institutions contributing to RACG research. Among universities, representative contributors include \texttt{Peking University}, \texttt{Zhejiang University}, \texttt{University of Chinese Academy of Sciences}, \texttt{Carnegie Mellon University}, \texttt{Fudan University}, and \texttt{University of Illinois Urbana-Champaign}. Among industry labs, major contributors include \texttt{Microsoft}, \texttt{ByteDance}, \texttt{Amazon}, \texttt{Alibaba}, \texttt{Ant Group}, and \texttt{Salesforce}. The distribution indicates that RACG has attracted sustained attention from both academia and industry, reflecting its relevance to practical code intelligence systems.

\subsection{Manuscript Organization}

\begin{figure}[t]

    \centering

    \includegraphics[width=0.82\linewidth]{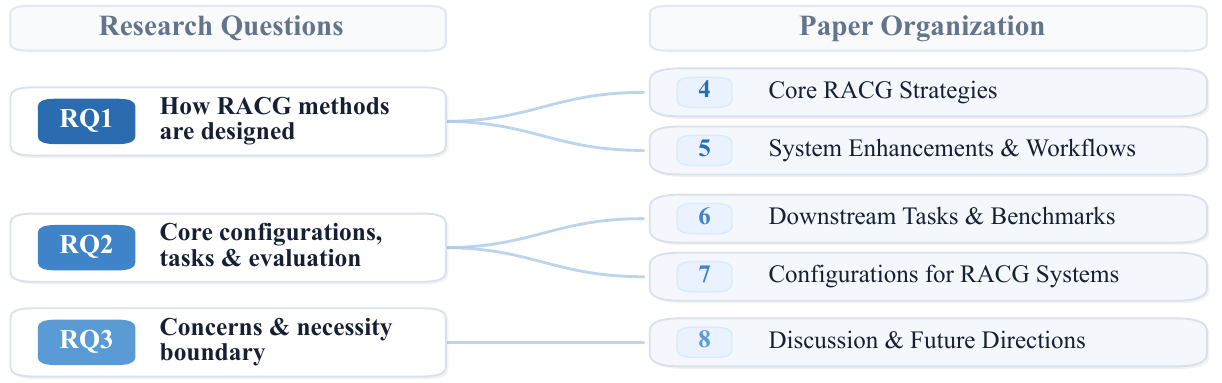}

    \caption{Mapping between the research questions and the organization of this survey.}

    \label{fig:rq-mapping}

\end{figure}

This section constitutes the main analytical body of our survey, synthesizing 152 selected papers on Retrieval-Augmented Code Generation with a particular emphasis on repository-level settings. To distinguish our work from prior general-purpose RAG surveys~\citep{surveycodegeneration,surveycodeagent}, we organize our findings around three coupled analytical dimensions: (1) \textbf{retrieval substrate} (\emph{how} context is represented and retrieved), (2) \textbf{control regime} (\emph{when} and \emph{how often} retrieval is invoked), and (3) \textbf{evaluation setting} (\emph{what tasks and benchmarks} reflect real-world demands). 
As illustrated in Figure~\ref{fig:rq-mapping}, these dimensions are reflected in the survey structure: Sections~\ref{RAG Strategies}–\ref{sec: enhancement} focus on retrieval substrate and control regime, while Sections~\ref{Downstream Tasks and Benchmarks}–\ref{sec:core_configurations} extend the discussion to downstream tasks, benchmarks, backbone models, and system configurations commonly adopted in RACG research. Section~\ref{sec:discussion} further provides a cross-cutting analysis of system-level limitations and the necessity boundary of RACG.
These dimensions also guide our three research questions. \textbf{RQ1} is addressed in Sections~\ref{RAG Strategies}–\ref{sec: enhancement}, \textbf{RQ2} in Sections~\ref{Downstream Tasks and Benchmarks}–\ref{sec:core_configurations}, and \textbf{RQ3} (core limitations and necessity boundary of RACG) in Section~\ref{sec:discussion}.
\section{Core RACG Strategies}
\label{RAG Strategies}
\definecolor{line-color}{RGB}{0, 128, 240}
\definecolor{fill-color}{RGB}{190, 225, 240}
\tikzstyle{category}=[
    rectangle,
    draw=line-color,
    rounded corners,
    text opacity=1,
    minimum height=1.5em,
    minimum width=5em,
    inner sep=2pt,
    align=center,
    fill opacity=.5,
]

\tikzstyle{leaf}=[category,minimum height=1em,
fill=fill-color!40, text width=20em,  text=black,align=left,font=\footnotesize,
inner xsep=4pt,
inner ysep=2pt,
]

\begin{figure*}[tp]
\centering
\scalebox{0.88}{
\begin{forest}
  forked edges,
  for tree={
  grow=east,
  reversed=true,
  anchor=base west,
  parent anchor=east,
  child anchor=west,
  base=left,
  font=\footnotesize,
  rectangle,
  draw=line-color,
  rounded corners,
  align=left,
  minimum width=3em,
  s sep=5pt,
  inner xsep=4pt,
  inner ysep=2pt,
  ver/.style={rotate=90, child anchor=north, parent anchor=south, anchor=center},
  },
  where level=1{text width=7em,font=\footnotesize,}{},
  where level=2{text width=7em,font=\footnotesize}{},
  where level=3{text width=8em,font=\footnotesize}{},
  where level=4{text width=17em,font=\footnotesize}{},
[RAG Strategies,ver
  [Non-Graph-based,ver
    [Retrieval Content \\ Construction
        [Query Reformulation \\ \& Multi-Perspective \\ Retrieval
            [ProCC~\citeyear{procc}{,}
             RLPG~\citeyear{RLPG}{,}
             ReCo~\citeyear{li-etal-2024-rewriting}{,}
             \\
             AlignCoder~\citeyear{jiang2026aligncoderaligningretrievaltarget}{,}
             SACL~\citeyear{gupta2025saclunderstandingcombatingtextual}{,}
             Li et al.~\citeyear{li2026contextintentreasoningguidedfunctionlevel}
             \\
             Kondo et al.~\citeyear{kondo-etal-2024-improving}{,}
             Code2JSON~\citeyear{singhal2025codejson}{,}
             \\
             Chen et al.~\citeyear{10.1145/3597503.3639085}{,}
             CodeBridge~\citeyear{10.1145/3729357}{,}
             ,leaf,text width=19em]
        ]
        [Context Structuring \\ \& Hierarchical \\ Construction
            [cAST~\citeyear{zhang2025castenhancingcoderetrievalaugmented}{,}
             NEMOTRON-CORTEXA~\citeyear{sohrabizadeh2025nemotroncortexa}{,}
             \\
             R$^2$C$^2$-Coder~\citeyear{r2c2coder}{,}
             A$^3$-CodGen~\citeyear{A^3-CodGen}{,}
            Hydra~\citeyear{leanh2026treatcodenaturallanguage}{,}
             \\
             RepoFuse~\citeyear{repofuse}{,}
             RAMBO~\citeyear{rambo}{,}
             ReCode~\citeyear{10.1145/3746252.3761035}{,}
             \\
            RepoGenix~\citeyear{10.1145/3691620.3695331}{,}
LongCodeZip~\citeyear{bogomolov2024longcodearenaset}{,}
             \\ RepoLens~\citeyear{wang2025extractingconceptualknowledgelocate}{,}
             CodeMem~\citeyear{wang2026codememastguidedadaptivememory}
             ,leaf,text width=19em]
        ]
        [Knowledge Base \\ Expansion \& \\ API Integration
            [EVOR~\citeyear{evor}{,}
             AllianceCoder~\citeyear{alliancecoder}{,}
             Deng et al.~\citeyear{11096713}
             ,leaf,text width=19em]
        ]
    ]
    [Retrieval Strategy \\ Optimization
        [Foundational Hybrid \\ Retrieval 
            [ReACC~\citeyear{reacc}{,}
             CEDAR~\citeyear{10172590}{,}
             RAP-Gen~\citeyear{10.1145/3611643.3616256}{,}
             \\
             CODEGENAPI~\citeyear{ZAN2025113934}
             ,leaf,text width=19em]
        ]
        [Adaptive Retrieval \\ \& Policy Learning
            [RepoCoder~\citeyear{repocoder}{,}
             kNN-TRANX~\citeyear{zhang-etal-2023-syntax}{,}
             \\
             SRACG~\citeyear{Wang_Ma_Gong_Wang_Chen_Cao_Cai_2026}{,}
             FT2Ra~\citeyear{10.1145/3650212.3652130}{,}ARCS~\citeyear{bhattarai2025arcsagenticretrievalaugmentedcode}{,}
             \\
             CODEFILTER~\citeyear{li2025impactdriven}{,}
             SWE-Fixer~\citeyear{SWE-Fixer}{,}
             \\
             ProCC~\citeyear{procc}{,}
             CodeRAG~\citeyear{zhang-etal-2025-coderag}{,}
             kNM-LM~\citeyear{10298575}
             ,leaf,text width=19em]
        ]
        [Scalable \& \\ Information-Efficient \\ Context Selection
            [RepoMinCoder~\citeyear{repomincoder}{,}
             RepoShapley~\citeyear{huo2026reposhapleyshapleyenhancedcontextfiltering}{,}
             \\
             HCP~\citeyear{HCP}{,}FastCoder~\citeyear{zhao2025fastcoderacceleratingrepositorylevelcode}{,}
             SpecAgent~\citeyear{ma2026specagentspeculativeretrievalforecasting}
             ,leaf,text width=19em]
        ]
        [Retrieval Fidelity \\ \& Structural \\ Awareness
            [CoRet~\citeyear{CoRet}{,}
             CCCI~\citeyear{jin2025cccicodecompletioncontextual}{,}
             HyRACC~\citeyear{11052791}{,}
             \\
             De-Hallucinator~\citeyear{eghbali2024dehallucinatormitigatingllmhallucinations}{,}
             Fedrushkov et al.~\citeyear{10.1007/978-3-031-97635-3_52}{,}
             \\
             Hydra~\citeyear{leanh2026treatcodenaturallanguage}
             ,leaf,text width=19em]
        ]
        [Learning from \\ Feedback \\ \& Self-Expression
            [RepoGenReflex~\citeyear{repogenreflex}{,}
             SelfRACG~\citeyear{dong2025selfracgenablingllmsselfexpress}
             ,leaf,text width=19em]
        ]
        [Low-Resource \\ Retrieval
            [RAR~\citeyear{dutta-etal-2024-rar}{,}
             PERC~\citeyear{yoo-etal-2025-perc}
             ,leaf,text width=19em]
        ]
    ]
  ]
  [Graph-based,ver,text width=5em
    [Retrieval Strategy \\ Optimization
        [Expansion Strategies \\ \& Traversal \\ Algorithms
            [RepoHYPER~\citeyear{RepoHyper}{,}
             RepoScope~\citeyear{liu2025enhancingrepositorylevelcodegeneration}{,}
             \\
             LingmaAgent~\citeyear{LingmaAgent}{,}
             SaraCoder~\citeyear{chen2025saracoderorchestratingsemanticstructural}{,}
             \\
             OrcaLoca~\citeyear{orcaloca}{,}
             RANGER~\citeyear{shah2025rangerrepositorylevelagent}{,}
             GRACE~\citeyear{wang2025gracegraphguidedrepositoryawarecode}{,}
             \\
             InlineCoder~\citeyear{hu2026linecontextrepositorylevelcode}{,}
             CodeGRAG~\citeyear{CodeGRAG}
             ,leaf,text width=19em]
        ]
        [External Signals \\ \& Structured \\ Databases
            [CocoGen~\citeyear{CocoGen}{,}
             AutoCodeRover~\citeyear{AutoCodeRover}{,}
             \\
             DSrepair~\citeyear{11029882}{,}
             Abedu~\citeyear{abedu2024synergizingllmsknowledgegraphs}
             ,leaf,text width=19em]
        ]
        [Graph-Based \\ Reasoning \\ Capabilities
            [SWE-Debate~\citeyear{li2025swedebatecompetitivemultiagentdebate}{,}
             Prometheus~\citeyear{chen2025prometheusunifiedknowledgegraphs}{,}
             \\
             GraphLocator~\citeyear{liu2025graphlocatorgraphguidedcausalreasoning}
             ,leaf,text width=19em]
        ]
    ]
    [Observed Design \\ Patterns
        [Foundational \\ Graph Structure
            [PKG~\citeyear{pkg}{,}
             GraphCoder~\citeyear{graphcoder}
             ,leaf,text width=19em]
        ]
        [Program Dependence \\ Modeling
            [DraCo~\citeyear{draco}{,}
             CodeGRAG~\citeyear{CodeGRAG}{,}
             GRACE~\citeyear{wang2025gracegraphguidedrepositoryawarecode}{,}
             \\
             GraphCoder~\citeyear{graphcoder}{,}
             SaraCoder~\citeyear{chen2025saracoderorchestratingsemanticstructural}
             ,leaf,text width=19em]
        ]
        [Knowledge Graph \\ Augmentation
            [ContextModule~\citeyear{ContextModule}{,}
             KGCompass~\citeyear{KGCompass}{,}
             \\
             PKG~\citeyear{pkg}{,}
             Abedu~\citeyear{abedu2024synergizingllmsknowledgegraphs}{,}
             Prometheus~\citeyear{chen2025prometheusunifiedknowledgegraphs}{,}
             \\
             SemanticForge~\citeyear{zhang2025semanticforgerepositorylevelcodegeneration}
             ,leaf,text width=19em]
        ]
        [Multi-Level \\ Graph Integration
            [CoSIL~\citeyear{CoSIL}{,}
             LocAgent~\citeyear{locagent}{,}
             RepoMaster~\citeyear{wang2026repomaster}{,}
             \\
             DraCo~\citeyear{draco}{,}
             CodeGRAG~\citeyear{CodeGRAG}{,}
             GRACE~\citeyear{wang2025gracegraphguidedrepositoryawarecode}{,}
             \\        GraphCodeAgent~\citeyear{li2025graphcodeagentdualgraphguidedllm}{,}GraphLocator~\citeyear{liu2025graphlocatorgraphguidedcausalreasoning}
             ,leaf,text width=19em]
        ]
        [Graph Integration \\ into Language \\ Models
            [CoCoMIC~\citeyear{cocomic}{,}
             CGM~\citeyear{cgm}{,}
             CodeRCSG~\citeyear{10967315}{,}
             \\
             SemanticForge~\citeyear{zhang2025semanticforgerepositorylevelcodegeneration}
             ,leaf,text width=19em]
        ]
    ]
  ]
  [New Trend,ver,text width=4em
    [Command-driven \\ Context Retrieval
      [ToolTrain~\citeyear{ma2025toolintegratedreinforcementlearningrepo}{,}
       RepoNavigator~\citeyear{zhang2026toolenoughreinforcementlearning}{,}
       CodeScout~\citeyear{sutawika2026codescouteffectiverecipereinforcement}{,}
       \\
       GrepRAG~\citeyear{wang2026grepragempiricalstudyoptimization}{,}
       SWE-Adept~\citeyear{he2026sweadeptllmbasedagenticframework}
       ,leaf,text width=28em]
    ]
    [Static Analysis \\ Integration
      [STALL+~\citeyear{stall}{,}
       MGD~\citeyear{MGD}{,}
       IDECoder~\citeyear{IDEcoder}{,}
       CatCoder~\citeyear{catcoder}{,}
       RRR~\citeyear{rrr}
       ,leaf,text width=28em]
    ]
  ]
]
\end{forest}
}
\caption{Taxonomy of RAG Strategies.}
\label{fig:taxonomyofRAG}
\end{figure*}
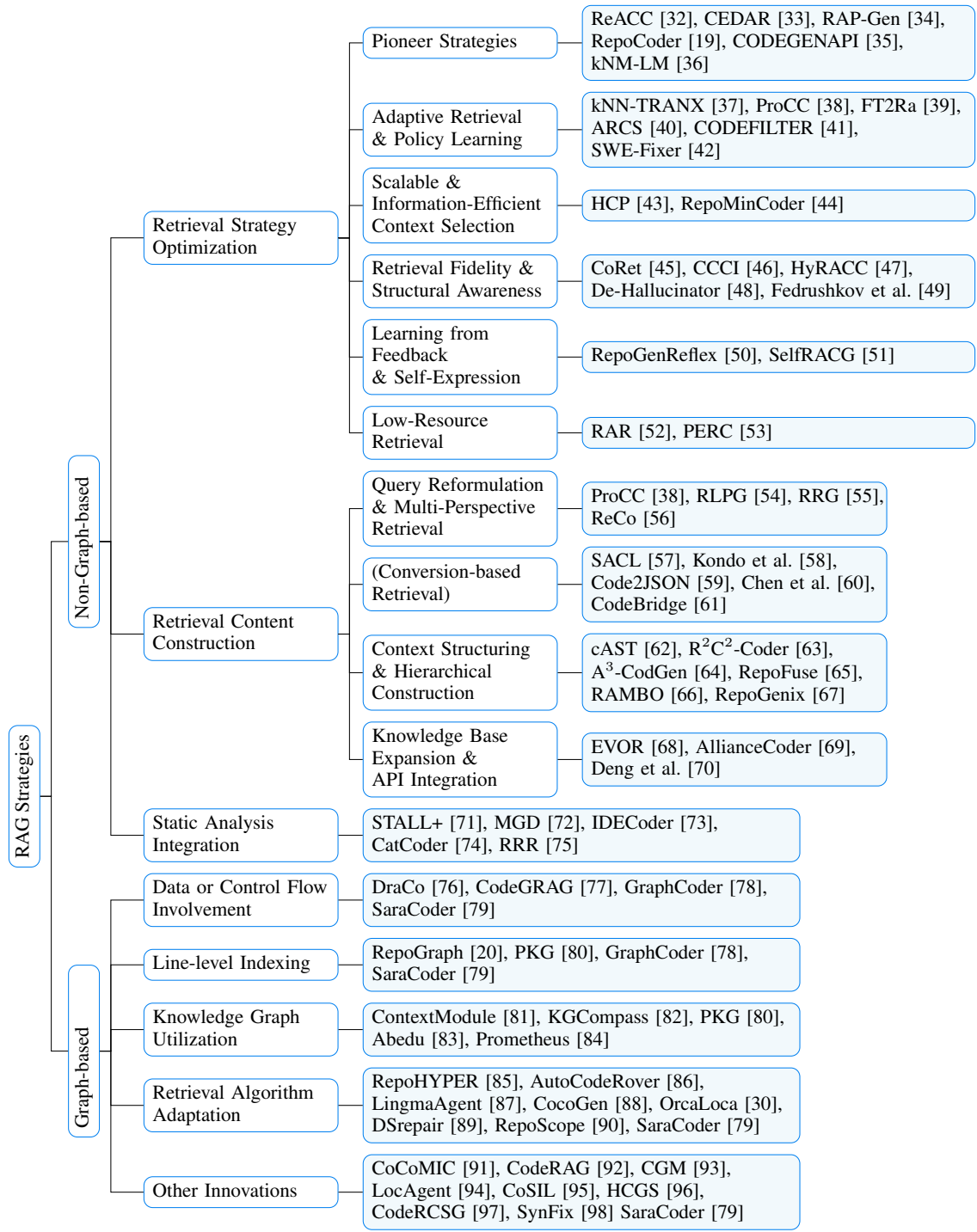

To systematically organize the RACG literature, we construct a taxonomy of RAG strategies through an iterative analysis of the selected studies. Specifically, we examine the retrieval substrate, context representation, and retrieval mechanisms adopted by prior work, and group methods according to their dominant retrieval characteristics. As illustrated in Figure~\ref{fig:taxonomyofRAG}, Retrieval-Augmented Code Generation methods can be broadly categorized into two major paradigms: \textbf{non-graph-based} and \textbf{graph-based}. We classify methods by their \emph{primary} retrieval substrate: a method is considered graph-based if its core retrieval mechanism depends on explicitly constructed graph structures, and non-graph-based otherwise. Methods that inject lightweight structural signals but retrieve over flat text collections remain in the non-graph category. Hybrid approaches are grouped according to their predominant component and noted explicitly. The two paradigms are detailed in turn below.

\subsection{Non-Graph-based RACG}
Non-graph-based RAG methods retrieve code without explicit graph structures, treating repositories as flat text collections and identifying relevant context via similarity matching. Early approaches rely on lexical signals (e.g., BM25~\citep{10.1007/978-1-4471-2099-5_24}, Jaccard similarity~\citep{jaccard1901}) to match identifiers or keywords, but struggle with deeper semantics.
Recent systems shift to semantic retrieval using representation learning. Models such as GraphCodeBERT~\citep{guo2021graphcodebert} and UniXcoder~\citep{guo-etal-2022-unixcoder} encode code into embeddings, enabling similarity matching that better captures functional and contextual relationships.

Despite their simplicity, these methods remain widely used due to efficiency and ease of integration. Modern RACG systems extend them along two axes: \textbf{retrieval strategies} (how to search) and \textbf{retrieval content construction} (what to retrieve and how to organize it). Emerging directions further incorporate \textbf{static analysis} for structural signals and \textbf{command-driven retrieval} for dynamic, interactive context acquisition.

\subsubsection{Retrieval Content Construction}

This line of work asks \textbf{\emph{what to retrieve}}. Rather than modifying retrieval algorithms, it improves performance by reformulating queries, restructuring content, and enriching the knowledge base.

\paragraph{Query Reformulation and Multi-Perspective Retrieval.}

A key issue is the mismatch between query form and code semantics. To address this, methods enhance query expressiveness in two ways.
First, \textit{multi-perspective enrichment} generates complementary query views. ProCC~\citep{procc} combines lexical, hypothesis, and summary views with bandit selection. RLPG~\citep{RLPG} samples candidate contexts and learns selection. ReCo~\citep{li-etal-2024-rewriting} rewrites queries and code for consistency. AlignCoder~\citep{jiang2026aligncoderaligningretrievaltarget} further leverages multiple candidate completions to construct richer queries.
Second, \textit{conversion-based retrieval} bridges modality gaps. SACL~\citep{gupta2025saclunderstandingcombatingtextual} augments code with semantic descriptions. Kondo et al.~\citep{kondo-etal-2024-improving} convert code to natural language. Code2JSON~\citep{singhal2025codejson}, Chen et al.~\citep{10.1145/3597503.3639085}, and CodeBridge~\citep{10.1145/3729357} explore structured or alternative matching. 
Li et al.~\citep{li2026contextintentreasoningguidedfunctionlevel} introduce reasoning-based intent synthesis via docstring-like generation.

\paragraph{Context Structuring and Hierarchical Construction.}

Another challenge is organizing retrieved content. Methods improve usability via structuring, hierarchy, and compression.
For \textit{structure-aware chunking}, cAST~\citep{zhang2025castenhancingcoderetrievalaugmented} uses AST-based partitioning, while Hydra~\citep{leanh2026treatcodenaturallanguage} preserves function-level integrity.
For \textit{multi-level organization}, R$^2$C$^2$-Coder~\citep{r2c2coder} builds abstract and fragment pools, A$^3$-CodGen~\citep{A^3-CodGen} integrates multi-granularity context, and NEMOTRON-CORTEXA~\citep{sohrabizadeh2025nemotroncortexa} emphasizes solution diversity.
For \textit{fusion and compression}, RepoFuse~\citep{repofuse}, RAMBO~\citep{rambo}, and RepoGenix~\citep{10.1145/3691620.3695331} compose compact prompts, while LongCodeZip~\citep{bogomolov2024longcodearenaset} applies coarse (function-level) and fine (block-level) pruning.
Search-space reduction is also explored: ReCode~\citep{10.1145/3746252.3761035} predicts task type, RepoLens~\citep{wang2025extractingconceptualknowledgelocate} clusters conceptual concerns, and CodeMem~\citep{wang2026codememastguidedadaptivememory} maintains API-level and session-level memory.

\paragraph{Knowledge Base Expansion and API Integration.}

When required knowledge is missing, the corpus itself must be expanded. EVOR~\citep{evor} dynamically updates both queries and the knowledge base for evolving domains. AllianceCoder~\citep{alliancecoder} generates API-level descriptions from queries, while Deng et al.~\citep{11096713} infer implicit API dependencies from drafts to retrieve relevant APIs.
Together, these methods improve retrieval by aligning queries with intent, structuring context for usability, and enriching the underlying knowledge base.

\subsubsection{Retrieval Strategy Optimization}

Retrieval strategy optimization studies \textbf{\emph{what} to fetch, \emph{when} to fetch it, and \emph{how much} to use}. These choices determine whether retrieved context helps or harms generation. Foundational work, like ReACC~\citep{reacc}, CEDAR~\citep{10172590}, and RAP-Gen~\citep{10.1145/3611643.3616256}, combine BM25 with dense embeddings, leveraging lexical matching for exact identifiers and dense retrieval for semantic similarity. CODEGENAPI~\citep{ZAN2025113934} extends this to private codebases via API signature retrieval.

\paragraph{Adaptive Retrieval and Policy Learning.}

Static strategies fail to adapt to query complexity and uncertainty. RepoCoder~\citep{repocoder} iteratively refines queries using intermediate generations, while kNM-LM~\citep{10298575} injects retrieval selectively for poorly predicted tokens.
Other methods learn adaptive policies. kNN-TRANX~\citep{zhang-etal-2023-syntax} applies syntax constraints with adaptive weighting. ProCC~\citep{procc} selects retrievers via LinUCB~\citep{linucb}, and FT2Ra~\citep{10.1145/3650212.3652130} uses $\Delta$logits to trigger retrieval. ARCS~\citep{bhattarai2025arcsagenticretrievalaugmentedcode} decomposes queries, while CODEFILTER~\citep{li2025impactdriven} and SWE-Fixer~\citep{SWE-Fixer} refine context usage.
CodeRAG~\citep{zhang-etal-2025-coderag} introduces log-probability-guided probing and multi-path retrieval. SRACG~\citep{Wang_Ma_Gong_Wang_Chen_Cao_Cai_2026} adds necessity-aware retrieval, multi-objective selection, and preference consistency filtering.

\paragraph{Scalable and Information-Efficient Context Selection.}

Large repositories exceed context limits, making selection critical. HCP~\citep{HCP} prunes context via structural reachability, while RepoMinCoder~\citep{repomincoder} optimizes information retention under length constraints. RepoShapley~\citep{huo2026reposhapleyshapleyenhancedcontextfiltering} models selection as a cooperative game using Shapley values.
Efficiency-oriented methods include FastCoder~\citep{zhao2025fastcoderacceleratingrepositorylevelcode}, which adopts speculative decoding with draft-and-verify stages, and SpecAgent~\citep{ma2026specagentspeculativeretrievalforecasting}, which precomputes speculative contexts via future-edit prediction.

\paragraph{Retrieval Fidelity and Structural Awareness.}

Reducing context size does not ensure correctness. CoRet~\citep{CoRet} incorporates call-graph structure, CCCI~\citep{jin2025cccicodecompletioncontextual} filters inconsistent files, and HyRACC~\citep{11052791} adaptively fuses model and datastore outputs.
To reduce hallucinations, De-Hallucinator~\citep{eghbali2024dehallucinatormitigatingllmhallucinations} verifies API usage, while Fedrushkov et al.~\citep{10.1007/978-3-031-97635-3_52} improve retriever robustness via syntactic hierarchy and contrastive learning. Hydra~\citep{leanh2026treatcodenaturallanguage} further learns dependency-aware retrieval through pairwise classification.

\paragraph{Learning from Feedback and Self-Expression.}

Retrieval is often decoupled from generation. RepoGenReflex~\citep{repogenreflex} introduces a Reflector that generates feedback to guide subsequent retrieval. SelfRACG~\citep{dong2025selfracgenablingllmsselfexpress} enables models to express information needs explicitly before retrieval.

\paragraph{Low-Resource Retrieval.}

In low-resource settings, weak corpora limit retrieval. RAR~\citep{dutta-etal-2024-rar} combines multiple weak sources (e.g., examples, grammar) via dual retrievers, while PERC~\citep{yoo-etal-2025-perc} converts tasks into pseudocode to enable cross-language retrieval.

\begin{table}[t]
\centering
\caption{Static analysis integration.}
\label{tab:static-analysis}
\begin{tabularx}{\linewidth}{@{}l>{\raggedright\arraybackslash}X@{}}
\toprule
\textbf{System} & \textbf{Usage} \\
\midrule
MGD~\citeyearpar{MGD} & Feedback loop ensures consistent generation of types, identifiers, API protocol, and enum values. \\
STALL+~\citeyearpar{stall} & Cross-file dependent contexts used in prompt formulation, decoding control, and post-processing. \\
IDECoder~\citeyearpar{IDEcoder} & Retrieves contexts of code elements, project structure, developer intention, and error feedback. \\
CatCoder~\citeyearpar{catcoder} & Builds a type dependency graph to guide prompts within scope for practicality. \\
RRR~\citeyearpar{rrr} & Iteratively refines context for undefined symbols and dependencies, class and member metadata. \\
\bottomrule
\end{tabularx}
\end{table}

\begin{trendminimal}[Emerging Direction: Static Analysis Integration]
A complementary direction incorporates \emph{static analysis} to provide explicit structural signals (e.g., types, dependencies, execution constraints), overcoming limitations of similarity-based retrieval. 

These signals can be integrated throughout the pipeline. Representative systems and their integration strategies are summarized in Table~\ref{tab:static-analysis}. STALL+~\citep{stall} applies static analysis in prompting, decoding, and post-processing. MGD~\citep{MGD} enforces type consistency via decoding-time masking. IDECoder~\citep{IDEcoder} leverages IDE-level analysis for precise context retrieval.
They also support prompt refinement. CatCoder~\citep{catcoder} uses type-dependency graphs to constrain prompts, while RRR~\citep{rrr} iteratively refines context with analysis and execution feedback.

Overall, static analysis improves robustness and correctness by adding explicit structural constraints.
\end{trendminimal}




\begin{trendminimal}[Emerging Direction: Command-driven Context Retrieval]
\label{Command-driven Context Retrieval}
A recent trend shifts retrieval from one-shot context selection to \emph{agent-driven search}, where LLMs actively decide what to query, which files or symbols to inspect, and when to stop.
ToolTrain~\citep{ma2025toolintegratedreinforcementlearningrepo}, RepoNavigator~\citep{zhang2026toolenoughreinforcementlearning}, and CodeScout~\citep{sutawika2026codescouteffectiverecipereinforcement} train agents to conduct multi-step repository exploration through tool calls or terminal commands. GrepRAG~\citep{wang2026grepragempiricalstudyoptimization} shows that LLM-generated \texttt{ripgrep} commands can serve as effective retrieval queries, while SWE-Adept~\citep{he2026sweadeptllmbasedagenticframework} uses agent-directed traversal to follow relevant dependency paths.

Overall, command-driven retrieval makes context acquisition more adaptive, reduces irrelevant context, and turns retrieval into a test-time decision process.
\end{trendminimal}




\begin{takeaway}[Takeaways]
The two dimensions correspond to core components of the RACG process. 
Retrieval strategy optimization controls the retriever $\mathcal{M}_r$ and its interaction with state $s_t$, enabling dynamic context selection. 
Retrieval content construction defines $\mathcal{C}(\mathcal{R}, s_t)$, determining how repository information is formatted into model inputs. 

Together, they form a unified design space balancing adaptability, informativeness, and correctness.
\end{takeaway}




\subsection{Graph-based RACG}

Graph-based RAG incorporates explicit structural information from code repositories by representing code as graphs rather than flat text. Although designs vary, they share the principle of leveraging structured representations to enhance retrieval. In practice, parsing tools such as Tree-sitter~\citep{Brunsfeld2013} are commonly used to extract Abstract Syntax Trees (ASTs), which form the basis for graph construction.

Formally, a repository is modeled as a graph $G = (V, E)$, where nodes $V$ represent code entities at different granularities, and edges $E$ encode structural, syntactic, or semantic relationships. This graph acts as a structured index, allowing retrieval to consider both textual similarity and relational context.

Table~\ref{tab:graph_based_methods} provides a systematic comparison of recent graph-based RACG approaches. When analyzing a graph-based system, two primary factors determine its effectiveness: \textbf{how the graph $G$ is constructed}, and \textbf{how retrieval is performed over the graph structure}. The following sections examine these two dimensions and summarize design patterns in current research.

\begin{table}[htbp]
  \centering
  \caption{Comparison of Graph-Based RAG Methods}
  \label{tab:graph_based_methods}
  \small 
  \setlength{\tabcolsep}{2pt} 
  \begin{tabularx}{\textwidth}{@{}l *{11}{C} @{}} 
    \toprule
    \multirow{2}{*}{\textbf{Method}} & 
    \multicolumn{6}{c}{\textbf{Edge Types}} & 
    \multicolumn{5}{c}{\textbf{Node Types}} \\
    \cmidrule(lr){2-7} \cmidrule(lr){8-12}
    & \textbf{Contain}
    & \textbf{Import}
    & \textbf{Inherit}
    & \textbf{Invoke}
    & \textbf{Data F.} 
    & \textbf{Ctrl F.} 
    & \textbf{Directory} 
    & \textbf{Module} 
    & \textbf{Class} 
    & \textbf{Function} 
    & \textbf{Line} \\
    \midrule
    AutoCodeRover~\citeyearpar{AutoCodeRover} & \checkmark & $\times$   & $\times$   & $\times$   & $\times$   & $\times$ & $\times$   & \checkmark & \checkmark & \checkmark & $\times$ \\
    CocoGen~\citeyearpar{CocoGen}     & \checkmark & \checkmark & \checkmark & $\times$   & $\times$   & $\times$ & $\times$   & \checkmark & \checkmark & \checkmark & $\times$ \\
    CoCoMIC~\citeyearpar{cocomic}     & \checkmark & \checkmark & $\times$   & $\times$   & $\times$   & $\times$ & $\times$   & \checkmark & \checkmark & \checkmark & $\times$ \\
    CodePlan~\citeyearpar{CodePlan}    & \checkmark & \checkmark & \checkmark & \checkmark & $\times$   & $\times$ & $\times$   & \checkmark & \checkmark & \checkmark & $\times$ \\
    ContextModule~\citeyearpar{ContextModule} & \checkmark & $\times$   & $\times$   & \checkmark & $\times$   & $\times$ & $\times$   & \checkmark & \checkmark & \checkmark & $\times$ \\
    DraCo~\citeyearpar{draco}       & \checkmark & \checkmark & \checkmark & \checkmark & \checkmark & $\times$ & \checkmark & \checkmark & \checkmark & \checkmark & $\times$ \\
    GraphCoder~\citeyearpar{graphcoder}  & $\times$   & $\times$   & $\times$   & \checkmark & \checkmark & \checkmark & $\times$   & $\times$   & $\times$   & $\times$   & \checkmark \\
    PKG~\citeyearpar{pkg} & \checkmark & $\times$   & $\times$   & $\times$   & $\times$   & $\times$ & $\times$   & $\times$   & $\times$   & \checkmark & \checkmark \\
    
    KG-RLCG~\citeyearpar{KG-RLCG}  & \checkmark & $\times$   & $\times$   & $\times$   & $\times$   & $\times$ & $\times$   & \checkmark & \checkmark & \checkmark & $\times$ \\
    CGM~\citeyearpar{cgm}         & \checkmark & \checkmark & \checkmark & \checkmark & $\times$   & $\times$ & \checkmark & \checkmark & \checkmark & \checkmark & $\times$ \\
    CodeGRAG~\citeyearpar{CodeGRAG}    & \checkmark & $\times$   & $\times$   & \checkmark & \checkmark & \checkmark & $\times$   & $\times$   & $\times$   & \checkmark & $\times$ \\
    GraphCodeAgent~\citeyearpar{li2025graphcodeagentdualgraphguidedllm}     & \checkmark & \checkmark & \checkmark & \checkmark & $\times$   & $\times$ & $\times$   & \checkmark & \checkmark & \checkmark & $\times$ \\
    CodexGraph~\citeyearpar{codexgraph}  & \checkmark & $\times$   & \checkmark & \checkmark & $\times$   & $\times$ & $\times$   & \checkmark & \checkmark & \checkmark & $\times$ \\
    CoSIL~\citeyearpar{CoSIL}        & \checkmark & \checkmark & \checkmark & \checkmark & $\times$   & $\times$ & \checkmark & \checkmark & \checkmark & \checkmark & $\times$ \\
    GRACE~\citeyearpar{wang2025gracegraphguidedrepositoryawarecode} & \checkmark & \checkmark & \checkmark & \checkmark & \checkmark & \checkmark & \checkmark & \checkmark & \checkmark & \checkmark & $\times$ \\
    GraphLocator~\citeyearpar{liu2025graphlocatorgraphguidedcausalreasoning} & \checkmark & \checkmark & \checkmark & \checkmark & $\times$ & $\times$ & \checkmark & \checkmark & \checkmark & \checkmark & $\times$ \\
    HCGS~\citeyearpar{sounthiraraj2025codecrafthierarchicalgraphbasedcode} & \checkmark & \checkmark & \checkmark & \checkmark & $\times$ & $\times$ & $\times$ & \checkmark & \checkmark & \checkmark & $\times$ \\
    KGCompass~\citeyearpar{KGCompass}   & \checkmark & $\times$   & $\times$   & \checkmark & $\times$   & $\times$ & $\times$   & \checkmark & \checkmark & \checkmark & $\times$ \\
    LingmaAgent~\citeyearpar{LingmaAgent} & \checkmark & $\times$   & \checkmark & \checkmark & $\times$   & $\times$ & \checkmark & \checkmark & \checkmark & \checkmark & $\times$ \\
    LocAgent~\citeyearpar{locagent}    & \checkmark & \checkmark & \checkmark & \checkmark & $\times$   & $\times$ & \checkmark & \checkmark & \checkmark & \checkmark & $\times$ \\
    OrcaLoca~\citeyearpar{orcaloca}    & \checkmark & $\times$   & $\times$   & \checkmark & $\times$   & $\times$ & \checkmark & \checkmark & \checkmark & \checkmark & $\times$ \\
    Prometheus~\citeyearpar{chen2025prometheusunifiedknowledgegraphs} & \checkmark & $\times$ & $\times$ & $\times$ & $\times$ & $\times$ & \checkmark & \checkmark & \checkmark & \checkmark & $\times$ \\
    RepoGraph~\citeyearpar{repograph}   & \checkmark & $\times$   & $\times$   & \checkmark & $\times$   & $\times$ & $\times$   & $\times$   & \checkmark & \checkmark & \checkmark \\
    RepoHYPER~\citeyearpar{RepoHyper}   & \checkmark & \checkmark & \checkmark & \checkmark & $\times$   & $\times$ & $\times$   & \checkmark & \checkmark & \checkmark & $\times$ \\
    
    RepoScope~\citeyearpar{liu2025enhancingrepositorylevelcodegeneration} & \checkmark & \checkmark & \checkmark & \checkmark & $\times$ & $\times$ & $\times$ & $\times$ & \checkmark & \checkmark & $\times$ \\
    SaraCoder~\citeyearpar{chen2025saracoderorchestratingsemanticstructural} & \checkmark & \checkmark & $\times$ & $\times$ & \checkmark & \checkmark & $\times$ & \checkmark & \checkmark & \checkmark & $\times$ \\
    SemanticForge~\citeyearpar{zhang2025semanticforgerepositorylevelcodegeneration} & \checkmark & \checkmark & \checkmark & \checkmark & $\times$ & $\times$ & $\times$ & \checkmark & \checkmark & \checkmark & $\times$ \\
    SWE-Debate~\citeyearpar{li2025swedebatecompetitivemultiagentdebate} & \checkmark & \checkmark & \checkmark & \checkmark & $\times$ & $\times$ & $\times$ & $\times$ & \checkmark & \checkmark & $\times$ \\
    SynFix~\citeyearpar{tang-etal-2025-synfix} & \checkmark & \checkmark & $\times$ & \checkmark & $\times$ & $\times$ & \checkmark & \checkmark & \checkmark & \checkmark & $\times$ \\

    InlineCoder~\citeyearpar{hu2026linecontextrepositorylevelcode} & $\times$ & $\times$ & $\times$ & \checkmark & $\times$ & $\times$ & $\times$ & $\times$ & $\times$ & \checkmark & $\times$ \\
    RepoMaster~\citeyearpar{wang2026repomaster} & \checkmark & \checkmark & $\times$ & \checkmark & $\times$ & $\times$ & \checkmark & \checkmark & \checkmark & \checkmark & $\times$ \\
    RPG~\citeyearpar{luo2026rpg} & \checkmark & \checkmark & $\times$ & \checkmark & \checkmark & $\times$ & \checkmark & \checkmark & \checkmark & \checkmark & $\times$ \\
    SWE-Adept~\citeyearpar{he2026sweadeptllmbasedagenticframework} & \checkmark & $\times$ & $\times$ & \checkmark & $\times$ & $\times$ & $\times$ & \checkmark & \checkmark & \checkmark & $\times$ \\
    \bottomrule
  \end{tabularx}
\end{table}

\subsubsection{Retrieval Content Construction}

\paragraph{Edge Types}  
Edges define typed relations between code entities. Each edge can be represented as a tuple $(v_i, v_j, r)$, where $v_i, v_j \in V$ and $r$ denotes the relation type. We consider six common types:

\begin{itemize}[leftmargin=2em]
  \item[\ding{217}] \textbf{Contain}: Captures structural inclusion, such as classes containing functions or modules containing classes. It may also represent hierarchical containment across multiple code blocks.
  
  \item[\ding{217}] \textbf{Import}: Encodes static dependencies introduced by language-level import or include statements across modules or packages.
  
  \item[\ding{217}] \textbf{Inherit}: Represents inheritance relationships between classes, reflecting object-oriented structure.
  
  \item[\ding{217}] \textbf{Invoke}: Captures call relations between functions or methods, modeling execution-level dependencies.
  
  \item[\ding{217}] \textbf{Data Flow (Data F.)}: Tracks how data propagates between variables, parameters, or return values.
  
  \item[\ding{217}] \textbf{Control Flow (Ctrl F.)}: Represents control dependencies induced by conditionals, loops, and branching constructs.
\end{itemize}

\paragraph{Node Types}  
Nodes correspond to code entities at different levels of granularity, forming a hierarchical representation of the repository:

\begin{itemize}[leftmargin=2em]
  \item[\ding{217}] \textbf{Directory}: Represents the top-level physical organization of a repository.
  
  \item[\ding{217}] \textbf{Module}: Corresponds to a single source file (e.g., \texttt{.py}, \texttt{.java}, \texttt{.cpp}). Although often referred to as \textit{File} in prior work, we use \textit{Module} to emphasize its role as a logical layer between Directory and Class.
  
  \item[\ding{217}] \textbf{Class} / \textbf{Function}: Core semantic units in most programming languages and common targets for retrieval.
  
  \item[\ding{217}] \textbf{Line}: The smallest addressable unit in code, enabling fine-grained retrieval and alignment.
\end{itemize}

\subsubsection{Retrieval Strategy Optimization}
Given a repository graph $G = (V, E)$, retrieval can be formulated as selecting a subgraph $G' = (V', E')$ conditioned on a query $q$. A common workflow first identifies seed nodes via lexical or semantic matching, then expands them to incorporate structural context. 

Expansion strategies determine how structure is utilized. Local methods such as $k$-NN or $n$-hop subgraphs capture nearby dependencies; Breadth-First Search emphasizes immediate neighbors, while Depth-First Search explores deeper paths. Some agent-based systems further define specialized primitives over ASTs (e.g., \texttt{search\_node}, \texttt{traverse\_graph}~\citep{locagent}). Pruning is typically applied to control context size.

\paragraph{Expansion Strategies and Traversal Algorithms.}
Existing methods can be viewed as different instantiations of expansion and traversal strategies, including redesigned relevance scoring and diverse search algorithms. RepoHYPER~\citep{RepoHyper} follows a search-expand-refine paradigm based on $k$-NN with a learned link predictor for re-ranking. RepoScope~\citep{liu2025enhancingrepositorylevelcodegeneration} leverages K-means to predict call chains. LingmaAgent~\citep{LingmaAgent} and RANGER~\citep{shah2025rangerrepositorylevelagent} cast traversal as a search problem using Monte Carlo Tree Search. OrcaLoca~\citep{orcaloca} lets LLM schedule retrieval order via relevance scores. SaraCoder~\citep{chen2025saracoderorchestratingsemanticstructural} improves retrieval diversity using Decaying Subgraph Edit Distance and Maximal Marginal Relevance. InlineCoder~\citep{hu2026linecontextrepositorylevelcode} retrieves callees after drafting code, while CodeGRAG~\citep{CodeGRAG} and GRACE~\citep{wang2025gracegraphguidedrepositoryawarecode} employ Graph Neural Networks to predict node relevance.

\paragraph{Integration of External Signals and Databases.}
Some approaches incorporate external signals or structured databases. CocoGen~\citep{CocoGen} refines retrieval using compiler feedback, and AutoCodeRover~\citep{AutoCodeRover} combines hierarchical search with fault localization. DSrepair~\citep{11029882} models code as RDF triples for SPARQL queries, while \citet{abedu2024synergizingllmsknowledgegraphs} translate natural language into Cypher queries.

\paragraph{Graph-Based Reasoning Capabilities.}
Other work enhances reasoning over graphs. SWE-Debate~\citep{li2025swedebatecompetitivemultiagentdebate} uses multi-agent debate over fault propagation paths. Prometheus~\citep{chen2025prometheusunifiedknowledgegraphs} introduces a working memory to avoid redundant reasoning and traversal, and GraphLocator~\citep{liu2025graphlocatorgraphguidedcausalreasoning} constructs causal graphs for step-by-step root cause analysis.

\subsubsection{Observed Design Patterns}
\paragraph{Foundational Graph Structure.} 
Most methods adopt a three-tier abstraction of \texttt{Module}, \texttt{Class}, and \texttt{Function}, connected by \texttt{Contain} edges for hierarchy and \texttt{Invoke} edges for call relations. This design preserves structure while remaining tractable for language model reasoning. A few approaches use finer granularity for higher precision: PKG~\citep{pkg} employs block- and function-level units with pruning, while GraphCoder~\citep{graphcoder} builds statement-level graphs with control-flow and data-dependence edges, improving local accuracy at the cost of coverage and overhead.

\paragraph{Program Dependence Modeling.} 
Only a subset of methods explicitly model fine-grained dependencies. DraCo~\citep{draco} captures variable dependencies via extended dataflow analysis, while CodeGRAG~\citep{CodeGRAG}, GraphCoder~\citep{graphcoder}, GRACE~\citep{wang2025gracegraphguidedrepositoryawarecode}, and SaraCoder~\citep{chen2025saracoderorchestratingsemanticstructural} incorporate both control flow and data flow. Adoption remains limited due to reliance on language-specific analysis, cross-function complexity, and high construction cost, although such graphs provide the most precise semantics without execution.

\paragraph{Knowledge Graph Augmentation.} 
Some methods extend graphs with repository metadata and external context. ContextModule~\citep{ContextModule} integrates user behavior and code signals, KGCompass~\citep{KGCompass} links code with issues and pull requests for multi-hop reasoning, and PKG~\citep{pkg} unifies code and text into a programming knowledge graph with multi-granularity retrieval. Abedu~\citep{abedu2024synergizingllmsknowledgegraphs} supports question answering over repository graphs, while Prometheus~\citep{chen2025prometheusunifiedknowledgegraphs} augments knowledge graphs with working memory for multi-step reasoning. SemanticForge~\citep{zhang2025semanticforgerepositorylevelcodegeneration} further applies knowledge graphs to repository-level code generation with correctness guarantees.

\paragraph{Multi-Level Graph Integration.}
A clear trend is the integration of multiple structural levels within a single system, where different graph representations are assigned complementary roles for context resolution. 
CoSIL~\citep{CoSIL} adopts a two-stage retrieval pipeline, first locating relevant files via a coarse module-level call graph and then refining context using a function-level graph. 
LocAgent~\citep{locagent} organizes repositories into a hierarchical dependency graph exposed as an interactive reasoning tool. 
Beyond structural hierarchies, several frameworks bridge modalities: GraphCodeAgent~\citep{li2025graphcodeagentdualgraphguidedllm} links requirement-level semantics with code structure, while DraCo~\citep{draco} and CodeGRAG~\citep{CodeGRAG} combine structural topology with semantic data flow to improve retrieval effectiveness. 
For autonomous exploration, RepoMaster~\citep{wang2026repomaster} integrates a Hierarchical Code Tree, Function Call Graph, and Module Dependency Graph. 
GRACE~\citep{wang2025gracegraphguidedrepositoryawarecode} further advances this direction with a multi-level, multi-semantic graph and hierarchical fusion. 
Finally, GraphLocator~\citep{liu2025graphlocatorgraphguidedcausalreasoning} constructs a Causal Issue Graph over a Repository Dependency Fractal Structure Graph to explicitly model symptom-to-root-cause paths.

\paragraph{Graph Integration into Language Models.} 
Several works study how to inject graph information into language models. CoCoMIC~\citep{cocomic} encodes code entities as tokens via causal attention, CGM~\citep{cgm} integrates graph structure into attention with dedicated adapters, and CodeRCSG~\citep{10967315} aligns cross-lingual semantic graphs with model representations. SemanticForge~\citep{zhang2025semanticforgerepositorylevelcodegeneration} incorporates graph constraints into beam search with formal verification.

\begin{takeaway}[Takeaways]
Taken together, graph-based RACG can be understood as operating over a structured retrieval space, where graph construction defines the search space, retrieval strategies determine how this space is explored, and integration mechanisms control how structural information influences generation.

Despite these advances, graph-based methods face a fundamental limitation: their construction pipelines depend on language-specific parsers and static analyzers, which reduces portability across programming languages. 
\end{takeaway}
\section{System Enhancements \& Workflows}
\label{sec: enhancement}

This section explores the methodologies for optimizing the end-to-end RACG pipeline. While modular RAG can be implemented in a lightweight, zero-shot manner, the inherent complexity often requires a tighter coupling between retrieval signals and the generation process. We categorize works into two primary paradigms:

First, \textbf{Training-Driven Optimizations} (Section \ref{Training-Driven Optimizations}) focus on parameter-level adaptations designed to align the retriever and generator modules. We primarily examine contrastive learning frameworks, architectural fusion techniques, and the emerging trend of using reinforcement learning.

Second, \textbf{Autonomous Agent Architectures} (Section \ref{sec:agentic_rag}) shift the paradigm from passive pipelines to dynamic workflows. To analyze these systems, we propose a three-tier classification framework based on autonomy. This allows RACG systems to refine decisions based on execution feedback.

\subsection{Training-Driven Optimizations}
\label{Training-Driven Optimizations}

RACG systems can be implemented in a lightweight manner, as work such as DraCo~\citep{draco} and RepoGraph~\citep{repograph} already showed promising results. Some research, however, increasingly treats training as essential for performance. Most approaches are self-supervised or unsupervised, leveraging the inherent structure of code in the absence of labeled data.

\subsubsection{Retrieval module training}

\paragraph{Contrastive learning and representation alignment.}
Most methods train retrievers to align queries and code in a shared space. ReACC~\citep{reacc} combines BM25 with a dense retriever and applies contrastive pre-training with code augmentations. InferFix~\citep{interfix} aligns buggy code with fixes, while CodeGenAPI~\citep{ZAN2025113934} pairs natural language queries with API documentation. CodeXEmbed~\citep{liu2025codexembed} unifies multiple tasks in a large-scale contrastive framework. SweRank~\citep{reddy2025sweranksoftwareissuelocalization} and SweRank+~\citep{wang2025extractingconceptualknowledgelocate} optimize retrieval with InfoNCE loss and hard negatives. NEMOTRON-CORTEXA~\citep{sohrabizadeh2025nemotroncortexa} generates synthetic query-code pairs, while CONAN-R~\citep{10.1145/3695868} enhances alignment with code--document and another mask entity prediction task.

\paragraph{Structure-aware optimization.}
Some methods incorporate code structure during training. kNN-TRANX~\citep{zhang-etal-2023-syntax} maps natural language to abstract syntax trees with auxiliary networks for reliability. CodeGRAG~\citep{CodeGRAG} jointly encodes textual and structural views using graph neural networks with structure-preserving learning.

\paragraph{Generation-driven and likelihood-based objectives.}
Other approaches optimize retrieval based on its impact on generation. CoRet~\citep{CoRet} introduces a likelihood-based loss using call graph and file path signals. CodeFilter~\citep{li2025impactdriven} scores retrieved blocks with special tokens. RepoShapley~\citep{huo2026reposhapleyshapleyenhancedcontextfiltering} models interactions among retrieved contexts via Shapley values. RLPG~\citep{RLPG} selects contexts with a trained classifier, while SelfRACG~\citep{dong2025selfracgenablingllmsselfexpress} integrates retrieval directly into the model’s hidden states.

\paragraph{Multi-stage pipelines and re-ranking.}
To balance scale and precision, several methods adopt coarse-to-fine retrieval. SWE-Fixer~\citep{SWE-Fixer} combines BM25 with a fine-tuned LLM model, SweRank~\citep{reddy2025sweranksoftwareissuelocalization} adds LLM re-ranking, and CodeRAG~\citep{zhang-etal-2025-coderag} distills re-rankers from stronger models.

\subsubsection{Generation module training}

\paragraph{Context augmentation and supervised fine-tuning.}
The most direct approach concatenates retrieved content with the input prompt.
ReACC~\citep{reacc} and InferFix~\citep{interfix} apply supervised fine-tuning (SFT) on augmented inputs, while R$^2$C$^2$-Coder~\citep{r2c2coder} and RepoFusion~\citep{repofusion} extend this idea by incorporating richer structural contexts and task alignment objectives.

\paragraph{Architectural fusion and attention mechanisms.}
Beyond prompt concatenation, several methods modify architectures for deeper integration.
Compression-based approaches such as CoCoMic~\citep{CocoGen}, CoRoVA~\citep{cherniuk2026corovacompressedrepresentationsvectoraugmented}, and HEF~\citep{sorokin2026hierarchicalembeddingfusionretrievalaugmented} reduce retrieved content into compact representations for efficient attention.
Graph-based methods, including CodeRCSG~\citep{10967315}, GRACE~\citep{wang2025gracegraphguidedrepositoryawarecode}, and CGM~\citep{cgm}, jointly model structural relationships via GNNs or subgraph pre-training.
Other designs, such as CONAN-G~\citep{10.1145/3695868}, leverage dual-view representations and FiD architectures to enhance semantic fusion.

\paragraph{Reasoning trajectories and distillation.}
To improve performance on complex tasks, some works incorporate reasoning supervision.
SWE-Fixer~\citep{SWE-Fixer} uses chain-of-thought (CoT) training for patch generation, while LocAgent~\citep{locagent} distills planning trajectories from larger models into smaller ones.
Other approaches explicitly model developer intent by generating intermediate representations (e.g., docstrings) to guide reasoning~\citep{li2026contextintentreasoningguidedfunctionlevel}.

\paragraph{Advanced training paradigms and domain adaptation.}
More advanced strategies refine training objectives and curricula.
RepoFormer~\citep{repoformer} learns when to do retrieval, while CMFT~\citep{CMFT} adopts curriculum learning for robustness.
SRI~\citep{zhang2026completioneditingunlockingcontextaware} reframes completion as dynamic editing, and domain-specific tuning is explored in DroidCoder~\citep{DroidCoder} and RTLRepoCoder~\citep{RTLRepoCoder}.

\begin{trendminimal}[Emerging Direction: Reinforcement learning for retrieval and generation alignment.]
To bridge the gap between retrieval and generation objectives, reinforcement learning (RL) is increasingly adopted.
Methods such as RLCoder~\citep{rlcoder} and AlignCoder~\citep{jiang2026aligncoderaligningretrievaltarget} optimize retrievers using generator feedback (e.g., perplexity), while SemanticForge~\citep{zhang2025semanticforgerepositorylevelcodegeneration} trains query planners with REINFORCE.
Two-stage pipelines combine SFT with preference optimization (e.g., PPO in RRG~\citep{RRG}, DPO in CoLA~\citep{10.1109/ASE63991.2025.00125}) to improve alignment and reduce hallucination.
RL also powers autonomous code agents such as RepoSearcher~\citep{ma2025toolintegratedreinforcementlearningrepo}, RepoNavigator~\citep{zhang2026toolenoughreinforcementlearning}, and CodeScout~\citep{sutawika2026codescouteffectiverecipereinforcement}.
Even outside RAG, works like SoRFT~\citep{sorft} and CoLT~\citep{li2025aixcoder7bv2trainingllmsfully} highlight RL's broader impact on improving code generation quality.
\end{trendminimal}

\begin{takeaway}[Takeaways]
Overall, training is crucial for RACG performance, but must align with task requirements and supervision. Key principles include:
\begin{itemize}[leftmargin=*]
  \item \textbf{Separate signals for retrieval and generation.} Fine-tuning the generator alone is often insufficient; retrieval-specific or joint objectives are needed when retrieval quality is the bottleneck.
  \item \textbf{Contrastive learning dominates retrieval training.} It effectively exploits code structure (e.g., transformations, bug--fix pairs) without costly annotation.
  \item \textbf{Reinforcement learning enables alignment.} It helps models learn when and how to use retrieved context, especially in multi-hop or ambiguous scenarios.
\end{itemize}
\end{takeaway}

\subsection{Autonomous Agent Architectures}
\label{sec:agentic_rag}

\begin{figure}[ht]
    \centering
    \begin{minipage}[t]{0.77\linewidth}
        \centering
        \includegraphics[width=\linewidth]{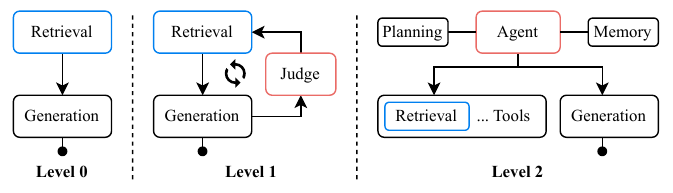}
    \end{minipage}%
    \hfill
    \begin{minipage}[t]{0.2\linewidth}
        \centering
        \includegraphics[width=\linewidth]{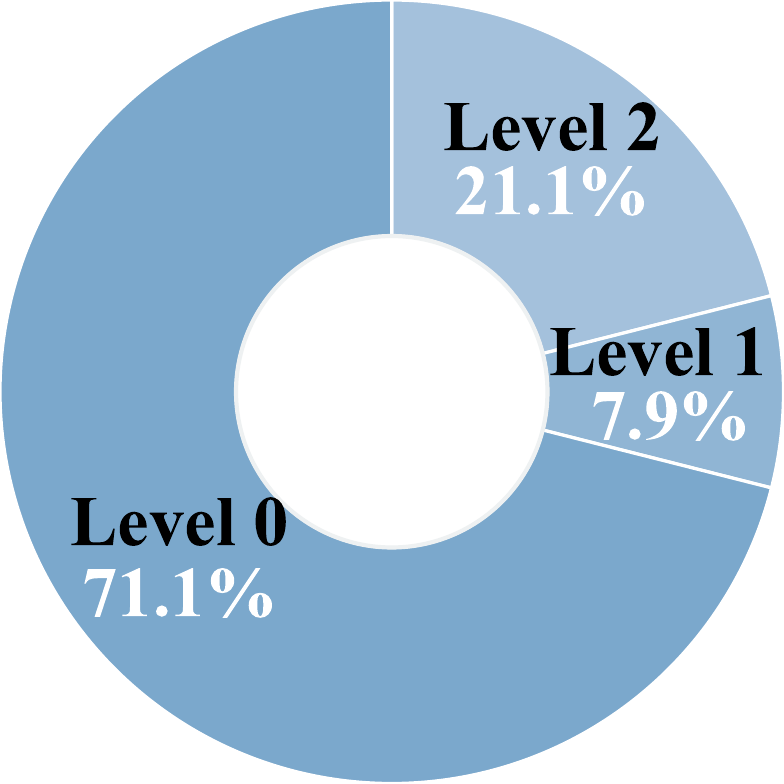}
    \end{minipage}
    \caption{Three-tier classification of agent-based architectures in RACG (left), ranging from non-agent systems (Level~0) to partial agent systems (Level~1) and fully autonomous agents (Level~2). The distribution of systems across levels is illustrated in the pie chart (right).}
    \label{fig:agent}
\end{figure}

Although recent RACG methods improve retrieval and context construction, many still operate as static, one-shot pipelines that cannot revise decisions based on intermediate outcomes. This limitation is especially evident in repository-level tasks (e.g., bug fixing and multi-file editing), which require iterative exploration, error correction, and interaction with external signals.

Agent-based architectures address this by enabling iterative decision-making, structured planning, and environment interaction, transforming RACG into dynamic, feedback-driven workflows. This allows models to decompose tasks, adapt retrieval and generation, and recover from errors.
Despite broad adoption across domains, the definition of an ``agent'' remains inconsistent. In this work, \textit{we define an agent as a system that perceives its environment, makes decisions via planning or learned policies, and iteratively executes actions to achieve goals without rigid control flow.}

To analyze agent-based RACG systems, we propose a three-tier taxonomy (Figure~\ref{fig:agent} and~\ref{fig:taxonomyofAgent}) based on autonomy:

\begin{itemize}[leftmargin=2em]
    \item \textbf{Level 0: Non-agent systems}—Static pipelines without iterative feedback, where retrieval and generation are performed in a single pass.
    
    \item \textbf{Level 1: Partial agent systems}—Systems with an explicit iterative feedback loop, where intermediate outputs (e.g., generated code, execution signals, or retrieved context) are used to refine subsequent decisions, but without autonomous planning or tool orchestration.
    
    \item \textbf{Level 2: Fully autonomous agents}—Systems that perform goal-directed reasoning with autonomous planning, multi-step decision-making, and interaction with external tools or environments.
\end{itemize}

\definecolor{line-color}{RGB}{0, 128, 240}
\definecolor{fill-color}{RGB}{190, 225, 240}
\tikzstyle{category}=[
    rectangle,
    draw=line-color,
    rounded corners,
    text opacity=1,
    minimum height=1.5em,
    minimum width=5em,
    inner sep=2pt,
    align=center,
    fill opacity=.5,
]

\tikzstyle{leaf}=[category,minimum height=1em,
fill=fill-color!40, text width=20em,  text=black,align=left,font=\footnotesize,
inner xsep=4pt,
inner ysep=2pt,
]

\begin{figure*}[tp]
\centering
\scalebox{0.9}{
\begin{forest}
  forked edges,
  for tree={
  grow=east,
  reversed=true,
  anchor=base west,
  parent anchor=east,
  child anchor=west,
  base=left,
  font=\footnotesize,
  rectangle,
  draw=line-color,
  rounded corners,
  align=left,
  minimum width=3em,
  s sep=5pt,
  inner xsep=4pt,
  inner ysep=2pt,
  ver/.style={rotate=90, child anchor=north, parent anchor=south, anchor=center},
  },
  where level=1{text width=7em,font=\footnotesize,}{},
  where level=2{text width=11em,font=\footnotesize}{},
  where level=3{text width=15em,font=\footnotesize}{},
  where level=4{text width=8em,font=\footnotesize}{},
[Agent Architectures,ver
    [L0: Non-agent,category
        [Most RACG work,leaf,text width=7em]
    ]
    [L1: Partial-agent,category
        [Architectural Interoperability
            [RepoGraph~\citeyearpar{repograph}{,}
             RepoLens~\citeyearpar{wang2025extractingconceptualknowledgelocate}
             ,leaf,text width=14.5em]
        ]
        [Self-guided Iteration
            [RepoCoder~\citeyearpar{repocoder}{,}
             FT2Ra~\citeyearpar{10.1145/3650212.3652130}{,}
             FastCoder~\citeyearpar{zhao2025fastcoderacceleratingrepositorylevelcode}{,} \\
             De-Hallucinator~\citeyearpar{eghbali2024dehallucinatormitigatingllmhallucinations}
             ,leaf,text width=21em]
        ]
        [Environment-driven Feedback
            [CocoGen~\citeyearpar{CocoGen}{,}
             EvoR~\citeyearpar{evor}{,}
             Tom et al.~\citeyearpar{Tony2025RetrieveRefineOrBoth}{,}
             Li et al.~\citeyearpar{li2026contextintentreasoningguidedfunctionlevel}
             ,leaf,text width=25em]
        ]
        [Coarse-to-fine Strategies
            [CoSIL~\citeyearpar{CoSIL}{,}
             APT~\citeyearpar{zhang2024llmbasedunittestgeneration}
             ,leaf,text width=11em]
        ]
    ]
    [L2: Fully-agent,category
        [Tool-augmented Systems,text width=10em
            [RepoNavigator~\citeyearpar{zhang2026toolenoughreinforcementlearning}{,}
             CodeScout~\citeyearpar{sutawika2026codescouteffectiverecipereinforcement}{,}
             GraphCodeAgent~\citeyearpar{li2025graphcodeagentdualgraphguidedllm}{,}
             \\
             NEMOTRON-CORTEXA~\citeyearpar{sohrabizadeh2025nemotroncortexa}{,}
             AutoCodeRover~\citeyearpar{AutoCodeRover}{,}
             RRR~\citeyearpar{rrr}{,}
             \\
             LocAgent~\citeyearpar{locagent}{,}
             RepoMaster~\citeyearpar{wang2026repomaster}{,}
             RTLFixer~\citeyearpar{10.1145/3649329.3657353}{,}
             \\
             iSWE~\citeyearpar{ganhotra2026resolvingjavacoderepository}{,}
             SWE-agent~\citeyearpar{sweagent}{,}
             OpenHands~\citeyearpar{openhands}
             ,leaf,text width=27em]
        ]
        [Multi-agent Coordination,text width=10em
            [CodexGraph~\citeyearpar{codexgraph}{,}
             SWE-Exp~\citeyearpar{chen2025sweexpexperiencedrivensoftwareissue}{,}
             SpecAgent~\citeyearpar{ma2026specagentspeculativeretrievalforecasting}{,}
             \\
             GraphLocator~\citeyearpar{liu2025graphlocatorgraphguidedcausalreasoning}{,}
             iSWE~\citeyearpar{ganhotra2026resolvingjavacoderepository}{,}
             SWE-Adept~\citeyearpar{he2026sweadeptllmbasedagenticframework}{,}
             \\
             SRI~\citeyearpar{zhang2026completioneditingunlockingcontextaware}{,}
             Prometheus~\citeyearpar{chen2025prometheusunifiedknowledgegraphs}{,}
             InfCode-C++~\citeyearpar{dong2025infcodecintentguidedsemanticretrieval}{,}
             \\
             SWE-Debate~\citeyearpar{li2025swedebatecompetitivemultiagentdebate}{,}
             HCAG~\citeyearpar{wu2026hcaghierarchicalabstractionretrievalaugmented}
             ,leaf,text width=22.5em]
        ]
        [Planning and \\ Search-based Exploration,text width=10em
            [RANGER~\citeyearpar{shah2025rangerrepositorylevelagent}{,}
             LingmaAgent~\citeyearpar{LingmaAgent}{,}
             OrcaLoca~\citeyearpar{orcaloca}{,}
             \\
             ARCS~\citeyearpar{bhattarai2025arcsagenticretrievalaugmentedcode}{,}
             SweRankAgent~\citeyearpar{wang2025extractingconceptualknowledgelocate}{,}
             ZeroRepo~\citeyearpar{luo2026rpg}
             ,leaf,text width=22em]
        ]
    ]
  ]
\end{forest}
}
\caption{Taxonomy of Agent Architectures.}
\label{fig:taxonomyofAgent}
\end{figure*}
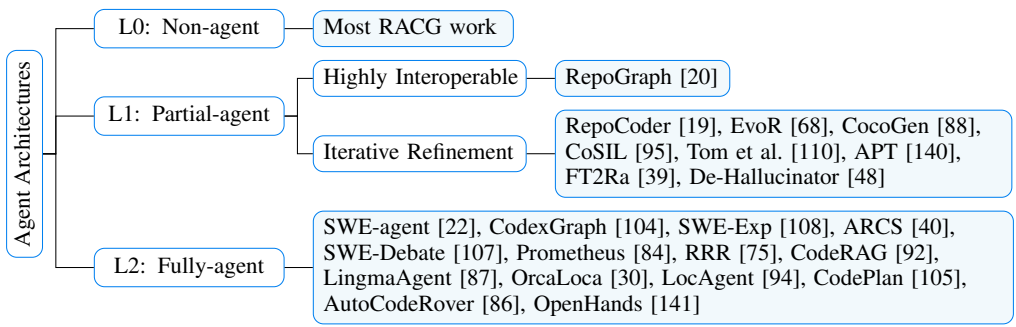

\subsubsection{Level 0: Non-agent Systems}

Most RACG approaches fall under Level 0, relying on fixed pipelines without dynamic decision-making. While effective for goal-driven tasks, this design lacks adaptivity for complex, multi-step repository problems.
Agentless~\citep{agentless} exemplifies this paradigm with a three-phase pipeline: hierarchical localization (repository $\rightarrow$ file $\rightarrow$ lines), patch generation, and regression-based validation. By avoiding agentic behavior, it achieves strong interpretability and low computational cost, while still reaching competitive performance (e.g., 32\% on SWE-bench Lite).

Notably, most training-intensive RACG systems also remain at this level, as retrievers and generators are often optimized independently and cannot adapt during inference, creating a gap between training-time optimization and runtime flexibility.

\subsubsection{Level 1: Partial Agent Systems}

Level 1 systems introduce iterative feedback loops, where intermediate signals (e.g., partial outputs, compiler feedback, or retrieved context) refine subsequent steps. Unlike Level 2, they do not require full planning or autonomous control. Architectural interoperability with agent frameworks, such as RepoGraph~\citep{repograph} and RepoLens~\citep{wang2025extractingconceptualknowledgelocate}, does not, by itself, constitute agentic behavior; such systems are therefore more appropriately classified as Level 1.

The other Level 1 systems can be grouped into three patterns. First, \textbf{self-guided iteration}, where model outputs guide retrieval, as in RepoCoder~\citep{repocoder}, FT2Ra~\citep{10.1145/3650212.3652130}, and FastCoder~\citep{zhao2025fastcoderacceleratingrepositorylevelcode}. De-Hallucinator~\citep{eghbali2024dehallucinatormitigatingllmhallucinations} further filters hallucinated APIs during iteration.
Second, \textbf{environment-driven feedback}, where external signals guide refinement. CocoGen~\citep{CocoGen} uses compiler feedback, EvoR~\citep{evor} incorporates execution and web signals, and other works refine outputs using domain constraints or human feedback~\citep{Tony2025RetrieveRefineOrBoth,li2026contextintentreasoningguidedfunctionlevel}.
Third, \textbf{coarse-to-fine strategies} progressively narrow the search space. CoSIL~\citep{CoSIL} moves from file-level to function-level reasoning, while APT~\citep{zhang2024llmbasedunittestgeneration} prioritizes simpler dependencies before complex ones.

A key challenge is stability: iterative loops may introduce error accumulation or semantic drift, requiring better control mechanisms or human-in-the-loop designs~\citep{repocoder}.

\subsubsection{Level 2: Fully Autonomous Agent Frameworks}

A growing number of RACG systems fall under Level 2, featuring fully agentic architectures with autonomous planning, tool usage, and multi-step reasoning. Unlike Level 1 systems that rely on predefined feedback loops, these systems act as goal-directed agents that dynamically explore and operate within repository environments. Existing approaches can be grouped into three main design patterns.

\paragraph{Tool-augmented systems.}
These systems rely on structured tools to navigate repositories, spanning multiple levels of abstraction. At the command level, RepoNavigator~\citep{zhang2026toolenoughreinforcementlearning} uses a \textit{jump} tool for symbol retrieval, while CodeScout~\citep{sutawika2026codescouteffectiverecipereinforcement} operates purely via Unix terminal commands. 
At the software engineering level, tools based on Abstract Syntax Trees (ASTs) and the Language Server Protocol (LSP) enable precise static analysis, as seen in RRR~\citep{rrr}, NEMOTRON-CORTEXA~\citep{sohrabizadeh2025nemotroncortexa}, and AutoCodeRover~\citep{AutoCodeRover}. 
Graph-based systems such as GraphCodeAgent~\citep{li2025graphcodeagentdualgraphguidedllm}, LocAgent~\citep{locagent}, and RepoMaster~\citep{wang2026repomaster} introduce traversal tools for repository graphs. Domain-specific frameworks include RTLFixer~\citep{10.1145/3649329.3657353} (Verilog compiler) and iSWE~\citep{ganhotra2026resolvingjavacoderepository} (CLDK for Java analysis). 
At the most general level, SWE-agent~\citep{sweagent} defines an Agent--Computer Interface, while OpenHands~\citep{openhands} provides a unified environment for command execution, browsing, and editing.

\paragraph{Multi-agent coordination.}
These systems decompose tasks across multiple specialized agents. Role-based designs separate responsibilities, e.g., CodexGraph~\citep{codexgraph} (natural language vs. Cypher query generation) and SWE-Exp~\citep{chen2025sweexpexperiencedrivensoftwareissue} (Instructor vs. Assistant). SpecAgent~\citep{ma2026specagentspeculativeretrievalforecasting} combines Retriever and Forecaster agents.
Task-level decomposition is also common, particularly separating localization and editing, as in GraphLocator~\citep{liu2025graphlocatorgraphguidedcausalreasoning}, iSWE~\citep{ganhotra2026resolvingjavacoderepository}, SWE-Adept~\citep{he2026sweadeptllmbasedagenticframework}, and SRI~\citep{zhang2026completioneditingunlockingcontextaware}. 
More complex pipelines include Prometheus~\citep{chen2025prometheusunifiedknowledgegraphs} (knowledge graph integration) and InfCode-C++~\citep{dong2025infcodecintentguidedsemanticretrieval} (Reproducer, Patch, Selector agents). 
Beyond pipelines, SWE-Debate~\citep{li2025swedebatecompetitivemultiagentdebate} introduces structured debate, while HCAG~\citep{wu2026hcaghierarchicalabstractionretrievalaugmented} employs cooperative game-inspired collaboration.

\paragraph{Planning and search-based exploration.}
These systems explicitly explore large search spaces. Tree-based methods such as RANGER~\citep{shah2025rangerrepositorylevelagent} and LingmaAgent~\citep{LingmaAgent} apply Monte Carlo Tree Search (MCTS). 
Efficiency-oriented designs include OrcaLoca~\citep{orcaloca} with action scheduling and context pruning and ARCS~\citep{bhattarai2025arcsagenticretrievalaugmentedcode} with plan-conditioned retrieval with tiered control. 
SweRankAgent~\citep{wang2025extractingconceptualknowledgelocate} structures exploration into search, reasoning, reformulation, and aggregation, while ZeroRepo~\citep{luo2026rpg} applies an explore--exploit strategy over a feature tree for scalable codebase generation.

A defining feature of Level 2 systems is their ability to enable open-ended exploration and adaptive decision-making. However, this flexibility introduces challenges in efficiency, controllability, and reliability at scale.

\begin{trendminimal}[Emerging Direction: Agentic Search Instead of Traditional RAG]
The shift from traditional RAG to Agentic Search reflects a move from passive similarity matching to active exploration. Claude Code~\citep{anthropic2025claudecode} exemplifies this paradigm by replacing static embeddings with dynamic tool usage. Instead of querying a fixed index, it leverages terminal utilities such as \texttt{grep}, \texttt{ls}, and \texttt{cat} within a continuous Plan--Act--Observe loop. This grounds reasoning in the live filesystem state, improving precision and enabling real-time dependency tracking. As discussed in Section~\ref{Command-driven Context Retrieval}, this aligns with command-driven context retrieval, and is increasingly adopted in systems such as CodeScout~\citep{sutawika2026codescouteffectiverecipereinforcement}.
\end{trendminimal}

\begin{takeaway}[Takeaways]
RACG agent architectures span a broad spectrum: Level~0 systems offer simplicity and strong baselines; Level~1 introduces iterative feedback for moderate adaptivity; and Level~2 enables full autonomy through planning, tools, and collaboration. Bridging the gap between Level~0 efficiency and Level~2 capability remains a key research challenge.
\end{takeaway}

\section{Downstream Tasks and Benchmarks for RLCG}
\label{Downstream Tasks and Benchmarks}

Since RACG is a broad topic, the benchmark landscape is highly fragmented, with many heterogeneous and loosely defined tasks. To ensure clarity and reliability, we focus only on formally defined benchmarks that are directly relevant to RLCG, centering on retrieval-augmented code generation rather than loosely related problems.

\subsection{Downstream Tasks}
\label{sec: tasks}
Before the emergence of RLCG, code intelligence research had already explored retrieval-centric tasks such as \textbf{code search}~\citep{yin2018mining}, which evaluates a model’s ability to retrieve or synthesize code from natural language queries, as well as \textbf{bug or vulnerability detection}~\citep{Li_2018}, which emphasizes identifying issues using retrieved evidence. With the rise of large language models, RLCG shifts the focus toward stronger \emph{generation} capabilities, integrating retrieval to support reasoning and synthesis rather than serving as the primary objective.

RLCG systems have since been applied to several key downstream tasks. The most prominent are \textbf{cross-file code completion} and \textbf{GitHub issue resolution}, both discussed in Section~\ref{rlcg-def}. These tasks reflect realistic software engineering scenarios that require reasoning beyond a single file: the former predicts missing code using repository context, while the latter automates bug fixing or feature implementation based on issue descriptions.

Another important setting is \textbf{coding ability evaluation}, represented by benchmarks such as HumanEval~\citep{humaneval} and MBPP~\citep{mbpp}. Although less grounded in real-world workflows, these benchmarks provide standardized environments to assess whether RLCG improves reasoning and code synthesis under external knowledge augmentation, and enable controlled comparisons with other techniques such as instruction tuning or chain-of-thought prompting.

Beyond these, RLCG has also been applied to \textbf{program repair}~\citep{Berabi2021TFixLT} and \textbf{test generation}~\citep{10.1145/3663529.3663839}, which focuses on synthesizing unit tests or assertions. Emerging directions further include \textbf{theory-to-code translation}~\citep{wu2026hcaghierarchicalabstractionretrievalaugmented} and \textbf{repository-level monitoring, editing, and refactoring}~\citep{CodePlan}. Together, these tasks highlight the versatility of RLCG across multiple dimensions of software engineering.

\subsection{Benchmarks}
We now turn to benchmarks for evaluating RLCG systems. Many works rely on self-constructed benchmarks, as they are easy to build and can highlight specific model strengths. However, this practice hinders fair and consistent comparison across systems.
To ensure rigor, we include only benchmarks adopted by multiple works and distinguish two roles:

\begin{itemize}[leftmargin=2em]
    \item \textbf{RLCG-direct benchmarks}: Designed for repository-level or cross-file generation, where retrieval over a codebase is intrinsic. These provide the primary evidence for evaluating RLCG systems.
    
    \item \textbf{General-purpose coding benchmarks}: Standalone tasks (e.g., algorithmic problems, docstring-to-code) that do not require repository context. These serve as \emph{auxiliary evidence} for assessing base LLM capability, but not RLCG effectiveness.
\end{itemize}

Representative benchmarks used in RLCG evaluation are summarized in Table~\ref{tab:code_benchmarks}.

\begin{table*}[ht]
\centering
\caption{Overview of representative benchmarks.}
\label{tab:code_benchmarks}
\renewcommand{\arraystretch}{1.15}
\setlength{\tabcolsep}{5pt}
\resizebox{\linewidth}{!}{

\begin{tabular}{@{}l l r l l l@{}}
\toprule
\textbf{Category} & \textbf{Benchmark} & \textbf{Size} & \textbf{Programming Language} & \textbf{Date} & \textbf{Link} \\
\midrule
\multirow{3}{*}{Line Completion} 
    & RepoEval~\citep{repocoder}        & 3573       & Python       & 2023-03 & \href{https://github.com/microsoft/CodeT/tree/main/RepoCoder}{link} \\
    & RepoBench~\citep{liu2024repobench}        &   49684\footnotemark[3]     & Python, Java       & 2024-01 & \href{https://github.com/Leolty/repobench}{link} \\
    & CrossCodeEval~\citep{crosscodeeval}        & 9928       & Python, Java, TypeScript, and C\#       & 2023-11 & \href{https://github.com/amazon-science/cceval}{link} \\
\midrule
\multirow{3}{*}{Function Generation} 
    & CoderEval~\citep{CoderEval}        &   460     & Python, Java      & 2024-02 & \href{https://github.com/CoderEval/CoderEval}{link} \\
    & DevEval~\citep{li-etal-2024-deveval}        &   1874     & Python      & 2024-05 & \href{https://github.com/seketeam/DevEval}{link} \\
    & EvoCodeBench~\citep{evocodebench}        &   275     & Python      & 2024-03 & \href{https://github.com/seketeam/EvoCodeBench}{link} \\
\midrule
\multirow{2}{*}{Real-World Resolution} 
    & SWE-bench~\citep{jimenez2024swebench}        &   2294     & Python      & 2023-10 & \href{https://www.swebench.com}{link} \\
    & Long Code Arena~\citep{bogomolov2024longcodearenaset}        &   ----\footnotemark[4]     & Python, Java, Kotlin       & 2024-06 & \href{https://huggingface.co/spaces/JetBrains-Research/long-code-arena}{link} \\
\midrule
\multirow{5}{*}{General Purpose} 
    & Aider Polyglot~\citep{aiderpolyglot} & 225 & Multi-language & 2024-12 & \href{https://aider.chat/docs/leaderboards/}{link} \\
    & LiveCodeBench~\citep{jain2025livecodebench} & 300+ & Python & 2024-03 & \href{https://livecodebench.github.io/}{link} \\
    & HumanEval~\citep{humaneval}        &   164     & Python      & 2021-07 & \href{https://huggingface.co/datasets/openai/openai_humaneval}{link} \\
    & MBPP~\citep{mbpp}        &   974     & Python      & 2021-08 & \href{https://github.com/google-research/google-research/tree/master/mbpp}{link} \\
    & CodeXGLUE~\citep{lu2021codexglue}        &   ----\footnotemark[5]     & Multi-language      & 2021-03 & \href{https://github.com/microsoft/CodeXGLUE}{link} \\
\bottomrule
\end{tabular}
}
\end{table*}

\footnotetext[3]{Calculated based on data from \href{https://huggingface.co/datasets/tianyang/repobench_python_v1.1}{RepoBench-Python} and \href{https://huggingface.co/datasets/tianyang/repobench_java_v1.1}{RepoBench-Java} on Hugging Face.}
\footnotetext[4]{Long Code Arena encompasses 6 distinct tasks, making it inappropriate to compute the overall size of the benchmark.}
\footnotetext[5]{CodeXGLUE encompasses 10 distinct tasks from 14 datasets, making it inappropriate to compute the overall size.}

\paragraph{Repository-Level Line Completion.}
\textbf{RepoEval}~\citep{repocoder} (from RepoCoder) includes 1,600 line completions, 1,600 API calls, and 373 function completions from high-quality Python repositories (2022+). It evaluates functional correctness using repository-native unit tests and snapshots (Jan 2023) across line, API, and function levels. 
\textbf{RepoBench}~\citep{liu2024repobench} supports Python and Java, decomposing RLCG into retrieval (RepoBench-R), completion (RepoBench-C), and pipeline (RepoBench-P) tasks. It uses newly crawled GitHub data and evaluates retrieval with \textit{Accuracy@k} and completion with \textit{EM/ES}. 
\textbf{CrossCodeEval}~\citep{crosscodeeval} provides 10,000 cross-file completion examples across Python, Java, TypeScript, and C\#, using static analysis for import resolution and evaluating in-file, cross-file retrieval, and combined prompt settings.

\paragraph{Repository-Level Function Generation.}
\textbf{CoderEval}~\citep{CoderEval} contains 460 real-world problems (Python/Java) across six context levels, focusing on non-standalone functions. It standardizes inputs as “signature + doc + context” and evaluates with Docker-based \texttt{Pass@k} and \texttt{Acc@k}. 
\textbf{EvoCodeBench}~\citep{evocodebench} offers 275 Python samples aligned with evolving repositories, incorporating dependency structures and updates; tasks combine natural language requirements, signatures, references, and tests, evaluated by \texttt{Pass@k} and \texttt{Recall@k}. 
\textbf{DevEval}~\citep{li-etal-2024-deveval} includes 1,874 manually annotated samples from 117 repositories across 10 domains, requiring function generation with signatures, natural language requirements, and cross-file context.

\paragraph{Pragmatic and Real-World Issue Resolution.}
\textbf{SWE-bench}~\citep{jimenez2024swebench} contains 2,294 real-world issues from 12 Python repositories, where models generate patches validated by tests; variants include SWE-bench-lite, -verified, -Pro, -multilingual, and -multimodal~\citep{yang2025swebench}. 
\textbf{Long Code Arena}~\citep{bogomolov2024longcodearenaset} consists of six long-context benchmarks for project-level tasks such as bug localization and library-based generation, with curated datasets and standardized evaluation tools.

\paragraph{General-Purpose Coding Benchmarks (Auxiliary Evidence).}
These benchmarks are not designed for repository-level or cross-file tasks but are often used to assess base LLM performance or compare augmentation strategies. They provide \emph{indirect evidence} for RACG effectiveness. 
\textbf{Polyglot Leaderboard}~\citep{aiderpolyglot} evaluates multilingual coding under a unified \texttt{pass@k} metric. 
\textbf{LiveCodeBench}~\citep{jain2025livecodebench} includes 300+ recent problems (2023--2024) for code generation, self-repair, and execution. 
\textbf{HumanEval}~\citep{humaneval} offers 164 Python problems with signatures, docstrings, and tests for \texttt{pass@k} evaluation. 
\textbf{MBPP}~\citep{mbpp} contains 974 beginner-level Python problems described in natural language. 
\textbf{CodeXGLUE}~\citep{lu2021codexglue} is a suite of 10 tasks across 14 datasets, covering translation, completion, and more, with unified evaluation and baselines.

\begin{takeaway}[Takeaways]
RLCG evaluation is shifting toward repository-level benchmarks that require integrating retrieval with generation (e.g., cross-file completion and issue resolution). General-purpose coding benchmarks offer only indirect evidence, making task-aligned, realistic evaluation increasingly important.
\end{takeaway}

\section{Configurations for RACG Systems}
\label{sec:core_configurations}

This section summarizes key configurations that shape RACG performance and scope. 
\textbf{Data Sources} (Section \ref{sec:data_sources}) highlight the shift from repository-local inputs to diverse, task-specific signals. 
\textbf{Programming Language Support} (Section \ref{sec:supported_languages}) reviews the Python-centric landscape and motivates cross-lingual frameworks for real-world use. 
\textbf{Backbone Models} (Section \ref{sec:backbone_models}) discusses model choices for retrieval and generation, along with their trade-offs.

\subsection{Data Source}
\label{sec:data_sources}

In RACG, retrieval typically operates over the target repository, sometimes augmented with API documentation~\citep{repocoder,draco,alliancecoder,11096713}. Recent work broadens this scope with task-specific resources, which can be grouped into four categories.

\textbf{General-purpose and web corpora.}
Some methods incorporate external programming-related text. PKG~\citep{pkg} uses Q\&A datasets and tutorials as general knowledge, while EvoR~\citep{evor} integrates heterogeneous web content (e.g., blogs, discussions, documentation, and execution feedback) into a unified, evolving knowledge base.

\textbf{Domain knowledge.}
Other approaches leverage specialized corpora or constraints. DeepV~\citep{ibnat2025deepv} reuses VerilogDB~\citep{calzada2025verilogdb} for hardware tasks. RTLFixer~\citep{10.1145/3649329.3657353} retrieves expert explanations for syntax errors, while CodeGuarder~\citep{lin2025llmssecuritycoursesecuring} and SecGuide-based methods~\citep{Tony2025RetrieveRefineOrBoth} incorporate vulnerability patterns and secure-coding rules.

\textbf{Repository-level structural and relational knowledge.}
Several systems exploit relations within or across repositories. A$^{3}$-CodGen~\citep{A^3-CodGen} links local context with global module structures and libraries. \citet{abedu2024synergizingllmsknowledgegraphs} construct repository knowledge graphs from commits, issues, and files. RepoMaster~\citep{wang2026repomaster} builds multi-level graphs (e.g., call graphs, dependency graphs) across repositories and retrieves task-relevant components.

\textbf{Experience and behavior traces.}
Another line leverages past interaction traces. ContextModule~\citep{ContextModule} mines real editing behavior, SWE-Exp~\citep{chen2025sweexpexperiencedrivensoftwareissue} stores repair trajectories, and ReCode~\citep{wang2025gracegraphguidedrepositoryawarecode} collects buggy--fixed pairs. Beyond code, \citet{ch-etal-2024-retrieval} use human interaction traces (e.g., collaborative building actions) as retrieval signals.

\begin{takeaway}[Takeaways]
RACG data sources are evolving from repository-local context to diverse external signals, including web corpora, domain knowledge, structural graphs, and behavioral traces. This diversification improves task performance but also highlights data selection as a key, under-explored design factor~\citep{wang-etal-2025-coderag}.
\end{takeaway}

\subsection{Programming Language Support}
\label{sec:supported_languages}

\begin{figure}
    \centering
    \includegraphics[width=0.7\linewidth]{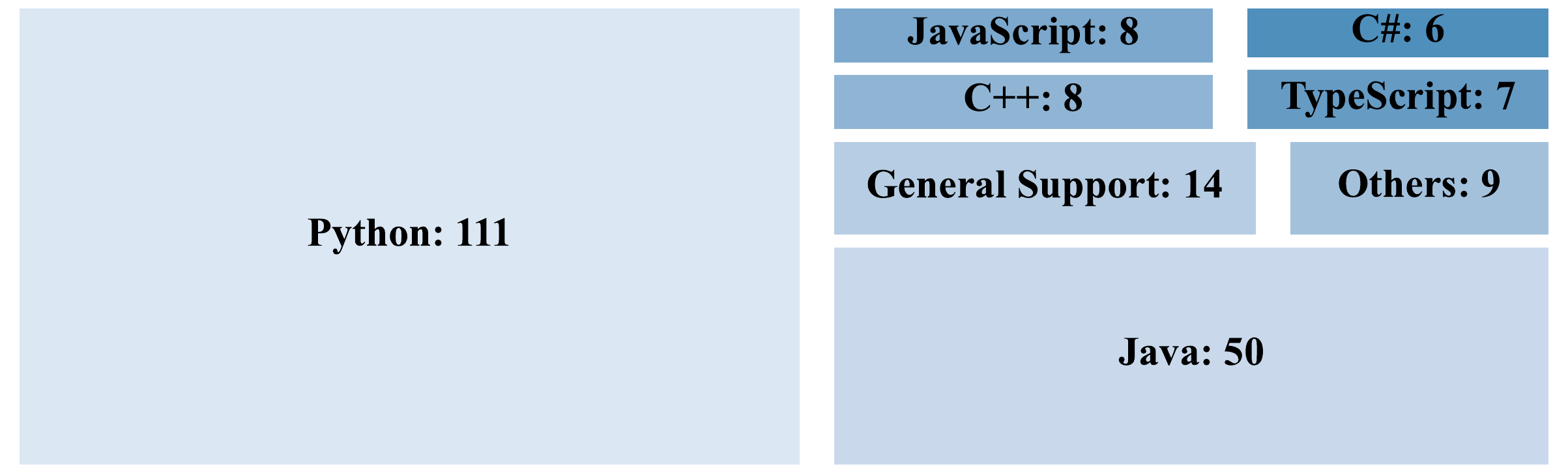}
    \caption{Programming Language Distribution}
    \label{fig:language_distribution}
\end{figure}

We analyze programming language usage in RACG systems based on explicit mentions in prior work, including \textit{cross-lingual transfer}.
As shown in Figure~\ref{fig:language_distribution}, \textbf{Python} dominates (111 occurrences), likely due to its prevalence in AI and data science, flexible syntax, and strong ecosystem. \textbf{Java} follows (50), reflecting its role in enterprise systems. \textbf{C++} and \textbf{JavaScript} appear less frequently (8 each), while \textbf{TypeScript} and \textbf{C\#} indicate continued interest in statically typed languages. Emerging languages such as \textbf{Rust}, \textbf{Go}, and \textbf{Kotlin} remain underrepresented.

This distribution shows a strong Python-centric bias in RACG research. While Python facilitates rapid prototyping, limited coverage of languages like Rust or Go may restrict generalization to real-world, multi-language codebases. Statically typed languages introduce additional challenges in structure, tokenization, and generation fidelity. As a result, bridging the gap between academic settings and heterogeneous industrial repositories remains an open problem, further compounded by benchmark bias (e.g., SWE-bench is Python-only).

\begin{trendminimal}[Emerging Direction: Universal Language Support]
Cross-language capability is increasingly important for RACG. \texttt{Zhu et al.}~\citep{zhu2025programminglanguagesilosstudy} show that cross-language retrieval can improve multilingual code generation, enhancing robustness and generalization.  

Many approaches map different languages into unified representations. \texttt{Code2JSON}~\citep{singhal2025codejson} converts code into structured natural language features via zero-shot LLMs. \texttt{CodeXEmbed}~\citep{liu2025codexembed} unifies retrieval tasks across text and code. \texttt{CodeRCSG}~\citep{10967315} builds cross-lingual semantic graphs using GNNs, while \texttt{Prometheus}~\citep{chen2025prometheusunifiedknowledgegraphs} constructs unified knowledge graphs across languages.  

Other works leverage the Language Server Protocol (LSP) for standardized multi-language support. \texttt{HCGS}~\citep{sounthiraraj2025codecrafthierarchicalgraphbasedcode} uses a modified \texttt{multilspy} client, and \texttt{ChatLSP}~\citep{10.1145/3689728} provides unified contextual services across languages.
\end{trendminimal}

\subsection{Backbone Models for Retrieval and Generation}
\label{sec:backbone_models}

We categorize and list the commonly used backbone models in RACG systems. 
We note that the models included here are representative rather than exhaustive, aiming to cover widely adopted examples across academic literature.

\paragraph{Dense Retrieval Models.}
Dense retrievers map queries and code into continuous vector spaces using neural encoders. \textbf{UniXcoder}~\citep{guo-etal-2022-unixcoder} leverages ASTs and comments with prefix-adapters and contrastive learning for accurate embeddings. \textbf{CodeT5}~\citep{codet5} incorporates identifier-aware pre-training to better capture semantic signals. \textbf{Voyage-Code}~\citep{voyagecode3} improves multilingual retrieval via Matryoshka learning and quantization-aware training. \textbf{Stella}~\citep{stella} uses multi-stage distillation for strong semantic representations. \textbf{GraphCodeBERT}~\citep{guo2021graphcodebert} models data flow dependencies beyond syntax, while \textbf{CodeBERT}~\citep{codebert} is effective for NL–PL tasks such as search and documentation. \textbf{Ada-Embedding-002}~\citep{openai2025textembeddingada002} provides general-purpose text–code embeddings with strong retrieval performance.

\paragraph{Sparse Retrieval Models.}
Sparse retrievers rely on lexical matching and simple similarity metrics. \textbf{BM25}~\citep{bm25} is a TF–IDF–based ranking function with length normalization, serving as a strong keyword-based baseline. \textbf{Jaccard Similarity}~\citep{jaccard1901} measures token overlap and is useful for capturing structural similarity between queries and code.

\begin{table}[ht]
\centering
\caption{Overview of selected open-source code generation models, including details such as the developing organization, model size (with some MoE models in the format of "full parameters (activation parameters)"), vocabulary size, context window length, total number of training tokens, and release date.}
\label{tab:encdec_llms}
\renewcommand{\arraystretch}{1.2}
\setlength{\tabcolsep}{4pt}
\resizebox{\linewidth}{!}{
\begin{tabular}{@{}l l l l c l c@{}}
\toprule
\textbf{Model} & \textbf{Organization} & \textbf{Size} & \textbf{Vocab} & \textbf{Context} & \textbf{Tokens} & \textbf{Date}\\
\midrule
CodeGen~\citep{codegen}      &  Salesforce AI    & 350M, 2B, 6B, 16B  & 50K   & 2048  & 577.2B   & 2022-03  \\ 
CodeGen2~\citep{codegen2}      &  Salesforce AI    & 1B, 3.7B, 7B, 16B  & 50K   & 2048  & 400B-1.4T   & 2023-03  \\ 
SantaCoder~\citep{santacoder}      &  BigCode    & 1.1B  & 48K   & 2048  & 236B   & 2023-01  \\ 
StarCoder(Base)~\citep{starcoder}      &  BigCode    & 1B, 3B, 7B, 15.5B  & 48K   & 8192  & 1T   & 2023-05  \\ 
StarCoder2~\citep{starcoder2}      &  BigCode    & 3B, 7B, 15B  & 48K   & 16K & 4T  & 2024-02  \\ 
Code LLaMA~\citep{codellama}      &  Meta    & 7B, 13B, 34B, 70B  & 31K   & 16K-100K & 500B-1T & 2023-08  \\ 
DeepSeek-Coder~\citep{deepseek-coder}      &  DeepSeek    & 1.3B, 5.7B, 6.7B, 33B  & 32K   & 16K & 2T  & 2024-01  \\ 
DeepSeek-Coder-V2~\citep{deepseekai2024deepseekcoderv2breakingbarrierclosedsource}      &  DeepSeek    & 16B(2.4B), 236B(21B)  & 100K   & 16K-128K & 6T  & 2024-06  \\ 
CodeQwen1.5~\citep{qwen}      &  Alibaba    & 7B  & 90K & 64K & 3T  & 2024-04  \\ 
Qwen2.5-Coder~\citep{qwen2coder}      &  Alibaba    &  0.5B, 1.5B, 3B, 7B, 14B, 32B  & 149K & 32K-128K & 5.5T  & 2024-11  \\ 
Qwen3-Coder~\citep{qwen3technicalreport}      &  Alibaba    & 30B(3B), 480B(35B)  & 148K & 256K-1M & 7.5T  & 2025-07  \\ 
\bottomrule
\end{tabular}
}
\end{table}

\paragraph{Open-source Generation Models.}
Representative open-source code generation models and their configurations are summarized in Table~\ref{tab:encdec_llms}. \textbf{CodeGen}~\citep{codegen} is a decoder-only Transformer family (up to $\sim$16B) trained on multilingual NL–PL data, showing strong zero-shot and multi-turn synthesis performance. 
\textbf{SantaCoder}~\citep{santacoder} is a compact 1.1B model trained on The Stack, leveraging Multi-Query Attention and Fill-in-the-Middle (FIM) for efficient inference and accurate infilling. 
\textbf{StarCoder}~\citep{starcoder} ($\sim$15.5B) is trained on $\sim$1T tokens across 80+ languages, supporting 8K context and achieving competitive HumanEval performance. 
\textbf{Code LLaMA}~\citep{codellama} extends LLaMA 2 for code (7B–34B), offering strong long-context completion and instruction-following. 
\textbf{DeepSeek-Coder}~\citep{deepseek-coder} is trained on $\sim$2T tokens (87\% code), with 16K context and strong multilingual completion and infilling. 
\textbf{Qwen2.5-Coder}~\citep{qwen2coder} supports up to 32B parameters, 128K context, and $\sim$92 languages, excelling in generation, repair, and reasoning.

\begin{table}[ht]
\centering
\caption{Overview of selected proprietary code generation models, including details such as the developing organization, context window length, cost, release date and official SWE-bench score.}
\label{tab:proprietary_llms}
\renewcommand{\arraystretch}{1.2}
\setlength{\tabcolsep}{4pt}
\begin{tabular}{@{}l l l l l l@{}}
\toprule
\textbf{Model} & \textbf{Organization} & \textbf{Context} & \textbf{Cost (Input/Output)} & \textbf{Date} & \textbf{SWE-bench}\\
\midrule
GPT-4o & OpenAI & 128K & \$2.50/\$10.00 & 2024-05 & 21.62\% \\
GPT-5 & OpenAI & 400K & \$1.25/\$10.00 & 2025-08 & 65.00\% \\
GPT-5 mini & OpenAI & 400K & \$0.25/\$2.00 & 2025-08 & 59.80\% \\
o3 & OpenAI & 200K & \$2.00/\$8.00 & 2025-04 & 58.40\% \\
Claude Opus 4  & Anthropic & 200K & \$15.00/\$75.00 & 2025-05 & 67.60\% \\
Claude Sonnet 4 & Anthropic & 200K & \$3.00/\$15.00 & 2025-05 & 64.93\% \\
Claude Sonnet 4.5 & Anthropic & 200K & \$3.00/\$15.00 & 2025-09 & 70.60\% \\
Gemini 2.5 Pro & Google & 1M & \$1.25/\$10.00 & 2025-03 & 53.60\% \\
Gemini 2.5 Flash & Google & 1M & \$0.30/\$2.50 & 2025-03 & 28.73\% \\
\bottomrule
\end{tabular}
\begin{tablenotes}
\small
\centering
\item * SWE-bench Verified score with mini-swe-agent
\item Costs shown are per million tokens (input/output) with no cache in USD
\item Context window shown as input token limit for Gemini Model(K = thousand, M = million)
\end{tablenotes}
\end{table}

\paragraph{Proprietary Generation Models.}
Representative proprietary code generation models and their capabilities are summarized in Table~\ref{tab:proprietary_llms}. \textbf{GPT and o Series (e.g., GPT-5, GPT-5 mini, GPT-4o, o3)}~\citep{openai2025gpt5} from OpenAI lead in code generation and reasoning, with GPT-5 targeting complex, long-context tasks and the o-series focusing on efficient multi-step reasoning. 
\textbf{Claude Series (e.g., Claude Opus 4, Sonnet 4.5)}~\citep{anthropic2025coding} from Anthropic excels in structured reasoning and coding, with Opus 4 supporting long-running agentic workflows and Sonnet 4 offering strong performance at lower cost. 
\textbf{Gemini Series (e.g., Gemini 2.5 Pro, Flash-Lite)}~\citep{deepmind2025gemini} from Google DeepMind provides strong coding and multimodal reasoning, with up to 1M-token context, while Flash-Lite emphasizes efficiency.


  
  

\begin{wrapfigure}{r}{0.4\textwidth}  
    \vspace{-3pt}  
    \centering
    \includegraphics[width=1.0\textwidth]{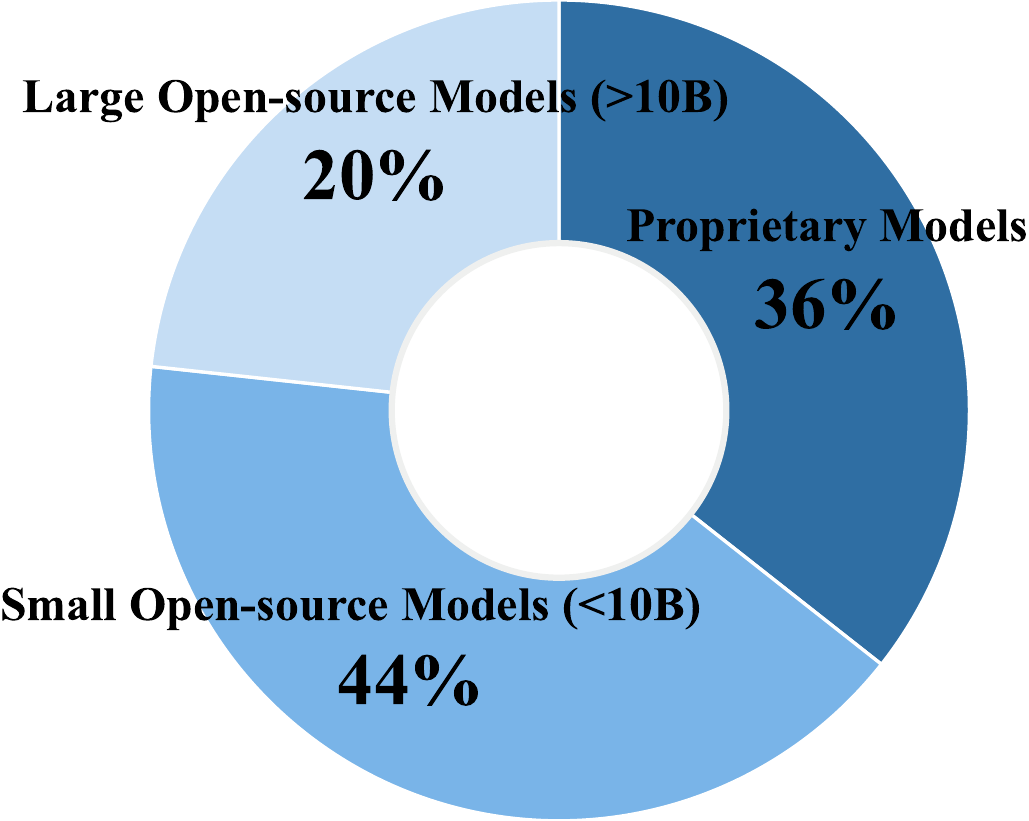}
    \caption{Distribution of base generation models used in the surveyed papers.}
    \label{fig:model_distribution}
    \vspace{-3pt}  
\end{wrapfigure}

\paragraph{Analysis.} We also conducted a statistical analysis of the base generation models adopted in each paper, focusing on the models used in the recommended configurations. This distinction allows us to better understand the models authors truly relied on in their proposed systems.
We categorized the models into three groups: (1) large open-source models (>10B parameters), (2) small open-source models (<10B parameters), and (3) proprietary models.

Figure~\ref{fig:model_distribution} reveals that small open-source models and proprietary models appear with comparable frequency and are significantly more popular than large open-source models. The prevalence of small models may be due to limited computational resources in academic or exploratory research settings, or because they can already achieve satisfactory performance on many tasks. Meanwhile, proprietary models are widely adopted, potentially due to two key reasons: (i) their superior performance on complex reasoning and generation tasks, and (ii) their ease of integration via API, reducing engineering and infrastructure overhead.
In contrast, large open-source models (>10B) are used far less frequently. This may stem from their high resource demands, including substantial GPU memory and inference time, which can pose barriers to adoption in both academic and industrial settings without dedicated infrastructure.

\begin{takeaway}[Takeaways]
Backbone selection in RACG involves key trade-offs:
\begin{itemize}[leftmargin=*]
  \item \textbf{Retrieval vs.\ context window:} Dense retrieval scales to large codebases, while long-context models simplify pipelines but incur higher cost and latency.
  \item \textbf{Sparse vs.\ dense retrieval:} Sparse methods (BM25) are efficient for keyword matching, whereas dense retrievers better capture semantics at higher cost.
  \item \textbf{Model size vs.\ deployment:} Smaller open models are accessible, while proprietary models lead performance; practical use depends on task complexity and system constraints.
\end{itemize}
\end{takeaway}

\section{Discussion and Future Directions}
\label{sec:discussion}

This section synthesizes the critical perspectives of RACG. As the field transitions from static augmentation to autonomous engineering, we analyze the critical boundaries of security, the necessity of retrieval in the era of expanding context windows, and the remaining bottlenecks that define future research opportunities.

\subsection{Security and Reliability}
The integration of RAG into repository-level development introduces a unique attack surface spanning both the generation and retrieval modules. On the generation side, LLMs risk inheriting insecure coding patterns from their training data, which often contains vulnerable open-source projects. On the retrieval side, knowledge base poisoning is a major concern. For instance, ImportSnare~\citep{10.1145/3719027.3765161} attack demonstrates that crafted malicious documentation can outrank safe resources, leading to the retrieval of harmful contexts \citep{10.1145/3719027.3765161}. DrainCode~\citep{wang2026draincodestealthyenergyconsumption} introduces LLM-Denial-of-Service, where the model is manipulated into generating excessively verbose code to exhaust computational resources. Furthermore, VenomRACG~\citep{li2025exploringsecuritythreatsretriever} demonstrates that a poisoned retriever can be trained to prioritize pre-defined vulnerable code when triggered by specific malicious queries.

To mitigate these risks, several safety-driven architectures have emerged. CodeGuarder~\citep{lin2025llmssecuritycoursesecuring} proposes that retrievers should also inject security knowledge. \cite{Tony2025RetrieveRefineOrBoth} combine security knowledge injection with a self-feedback loop.  RESCUE~\citep{shi2026rescue} introduces a systematic defense through a Hierarchical Security Knowledge Base and Hierarchical Multi-faceted Retrieval.

\subsection{The Necessity Boundary: RAG vs. Long-Context LLMs}

The effectiveness of RAG is context-dependent, and retrieval does not always guarantee better performance. \cite{peng2025can} show that long-context models can match or outperform RAG on small, well-structured repositories, while RAG becomes advantageous as repository size and complexity increase.

Several studies demonstrate the benefits of retrieval. CodeGen4Libs~\citep{10298327} validates the usefulness of retrieval for library-oriented generation through user studies. \cite{chen2025llmsmeetapidocumentation} show that RAG improves performance on uncommon APIs and remains robust to mild noise, with BM25 performing strongly. \cite{10.1145/3717061}, \cite{galimzyanov2025practicalcoderagscale} and \cite{hostnik2025128596} further report consistent gains from retrieval, especially when combined with effective fusion and selection strategies.

Other work explores replacing RAG with stronger long-context modeling. SelectSolve~\citep{selectsolve} shows that providing the full repository to a long-context model can rival or exceed agent-based systems. Additional approaches use hierarchical summarization, reinforcement learning, or tool-aware training (e.g., SoRFT~\citep{sorft}, CoLT~\citep{li2025aixcoder7bv2trainingllmsfully}, ToolGen~\citep{toolgen}) to improve context utilization without explicit retrieval.

Industry results also highlight trade-offs. Retrieval improves large-scale systems such as Tencent’s WeChat codebase~\citep{yang2025deepdiveretrievalaugmentedgeneration}, and combining RAG with fine-tuning yields further gains~\citep{10.1145/3696630.3728535}. However, fine-tuned models can better capture domain-specific conventions (“dialects”)~\citep{finkler2026automatedcustomizationllmsenterprise}, suggesting limits to retrieval alone.

\begin{takeaway}[Takeaways]
RAG is not universally superior. It excels in large or complex repositories, while long-context models can suffice in smaller settings. Effective systems often combine retrieval, model capacity, and fine-tuning, highlighting a key trade-off between context length, retrieval cost, and adaptability.
\end{takeaway}

\subsubsection{Limitations of Existing Approaches}
\label{Limitations of Existing Approaches}

Despite rapid progress, current RACG systems still face several important limitations in repository-level software engineering.

\textbf{Retrieval and context construction remain difficult.}
Non-graph-based approaches often retrieve fragmented repository chunks that fail to preserve long-range semantic dependencies, while graph-based systems introduce substantial overhead in graph construction and traversal~\citep{zhang2025castenhancingcoderetrievalaugmented,CodeGRAG,wang2025gracegraphguidedrepositoryawarecode}. Their effectiveness also heavily depends on language-specific parsers and static analysis quality, limiting portability across repositories and programming languages~\citep{stall,draco}. In addition, retrieval and generation are still frequently optimized separately, creating a disconnect between retrieved context quality and downstream generation utility~\citep{CoRet,dong2025selfracgenablingllmsselfexpress}.

\textbf{Agentic RACG systems often trade adaptivity for efficiency.}
Recent agent frameworks enable iterative search, planning, and tool usage, but they also introduce substantial context and computation overhead~\citep{sweagent,sutawika2026codescouteffectiverecipereinforcement}. Multi-step workflows may accumulate noisy context or redundant exploration steps as repositories scale~\citep{chen2025prometheusunifiedknowledgegraphs,shah2025rangerrepositorylevelagent}. At the same time, simpler pipelines such as Agentless can achieve competitive performance with significantly lower cost, suggesting that complex agentic workflows are not always necessary~\citep{agentless}.

\textbf{Current evaluation settings still lack realism and robustness.}
Most benchmarks focus on relatively small repositories, synthetic tasks, or Python-centric ecosystems, while repository evolution, collaborative workflows, and multi-language dependencies are rarely modeled explicitly~\citep{jimenez2024swebench}. Recent studies further show that static benchmarks may suffer from contamination and benchmark saturation~\citep{openai2026swebenchverified}. Moreover, most RACG systems are still evaluated in isolated environments rather than integrated software engineering ecosystems such as IDEs, CI/CD pipelines, and collaborative development workflows~\citep{sweagent,openhands}.

\subsubsection{Future Directions}
\label{Future Directions}

\textbf{RACG systems should move beyond text-only repositories toward richer software artifacts.}
Future systems can leverage issue discussions, documentation, execution logs, commit histories, and architectural diagrams to improve semantic grounding in underspecified or evolving tasks~\citep{KGCompass,ContextModule}. At the same time, multilingual repository support remains increasingly important, as modern software projects often combine multiple programming languages and frameworks within a single codebase~\citep{10967315,yoo-etal-2025-perc}.

\textbf{Scalable repository-level reasoning and realistic evaluation remain essential research directions.}
Recent work increasingly explores hierarchical retrieval, memory compression, and long-context reasoning to support repository-scale understanding with manageable computational cost~\citep{bogomolov2024longcodearenaset,huo2026reposhapleyshapleyenhancedcontextfiltering,ma2026specagentspeculativeretrievalforecasting}. Meanwhile, future benchmarks should better reflect practical software engineering workflows such as repository evolution, CI/CD integration, package migration, and collaborative development~\citep{li2026repolaunch}. Beyond pass@k or patch correctness, evaluation should also measure factors such as execution efficiency, integration consistency, maintainability, and developer utility~\citep{jimenez2024swebench}.

\textbf{Agentic software engineering workflows are becoming a central paradigm for RACG.}
Recent systems such as SWE-agent, OpenHands, Claude Code, and Codex demonstrate a shift from static retrieval-and-generation pipelines toward interactive software engineering agents capable of repository navigation, iterative refinement, and environment interaction~\citep{sweagent,openhands}. Retrieval is increasingly treated as an adaptive test-time process, where agents dynamically decide what files or resources to inspect during problem solving. Future RACG systems may further support repository-wide coordinated editing tasks requiring continuous reasoning over inter-file dependencies and evolving repository states~\citep{CodePlan,wang2026repomaster}.

\textbf{Bridging retrieval, reasoning, and generation remains the core challenge of RACG.}
Earlier code intelligence systems primarily focused on retrieval and matching, whereas modern LLM-based systems emphasize generative capability. Future RACG systems will likely move toward tighter integration between retrieval, memory, planning, reasoning, and generation, where retrieved repository artifacts dynamically guide and constrain generation decisions~\citep{dong2025selfracgenablingllmsselfexpress,zhang-etal-2025-coderag,chen2025prometheusunifiedknowledgegraphs}. Achieving this integration is essential for building software engineering agents that are both repository-grounded and capable of adaptive, context-aware reasoning.
\section{Conclusion}

As software engineering increasingly shifts from isolated code completion toward repository-scale reasoning and autonomous development workflows, Retrieval-Augmented Code Generation (RACG) has emerged as a critical paradigm for enabling large language models to operate effectively in real-world environments. Unlike traditional code generation settings, repository-level scenarios require models to continuously retrieve, organize, and reason over evolving contextual information distributed across files, dependencies, documentation, and execution feedback.

In this survey, we presented a comprehensive review of RACG research with a particular focus on Repository-Level Code Generation (RLCG). We argued that RACG should not be viewed as a static ``retrieve-then-generate'' pipeline, but rather as a coupled and dynamic process involving context construction, retrieval optimization, generation, and environment interaction. To systematically analyze this rapidly evolving landscape, we introduced a unified analytical framework spanning retrieval substrate, control regime, and evaluation setting. Based on this framework, we organized existing methods across non-graph-based and graph-based retrieval paradigms, training-driven optimizations, and autonomous agent architectures, while also reviewing downstream tasks, benchmarks, and practical system configurations.

Our analysis reveals several emerging trends in the field. First, RACG systems are evolving from similarity-based retrieval toward structure-aware and graph-driven reasoning. Second, retrieval is increasingly becoming an adaptive and agentic process rather than a fixed preprocessing step. Third, autonomous software engineering agents are transforming RACG from passive generation pipelines into interactive systems capable of planning, tool usage, iterative refinement, and environment-aware decision making. At the same time, important challenges remain unresolved, including scalability, retrieval reliability, long-context efficiency, benchmark realism, cross-language portability, and robustness in dynamic repositories.

Finally, we highlight the growing importance of understanding the necessity boundary between retrieval and long-context modeling. While long-context LLMs continue to improve rapidly, repository-level software engineering still fundamentally depends on efficient context selection, structured reasoning, and adaptive interaction with external environments. We believe future RACG systems will increasingly integrate retrieval, reasoning, planning, memory, and execution into unified agentic software engineering frameworks.

We hope this survey provides a structured foundation for future research and helps bridge the gap between academic advances in code generation and practical AI-powered software engineering systems.
\clearpage
\bibliography{ref}

\appendix
\clearpage

\end{document}